# Multiclass information flow propagation control under vehicle-to-vehicle communication environments

Jian Wang [a,b], Srinivas Peeta [c,1], Lili Lu [d], Tao Li [e]

[a] School of Civil Engineering, Purdue University, West Lafayette, IN 47907, United States
[b] NEXTRANS Center, Purdue University, West Lafayette, IN 47906, United States
[c] School of Civil and Environmental Engineering, and H. Milton Stewart School of Industrial and Systems Engineering, Georgia Institute of Technology, Atlanta, GA 30332, United States
[d] Faculty of Maritime and Transportation, Ningbo University, China
[e] Department of Computer Science, Purdue University, West Lafayette, IN 47907, United States

**Abstract:** Most existing models for information flow propagation in a vehicle-to-vehicle (V2V) communications environment are descriptive. They lack capabilities to control information flow, which may preclude their ability to meet application needs, including the need to propagate different information types simultaneously to different target locations within corresponding time delay bounds. This study proposes a queuing-based modeling approach to control the propagation of information flow of multiple classes. Two control parameters associated with a vehicle, the number of communication servers and the mean communication service rate, are leveraged to control the propagation performance of different information classes. A two-layer model is developed to characterize the information flow propagation wave (IFPW) under the designed queuing strategy. The upper layer is formulated as integro-differential equations to characterize the spatiotemporal information dissemination due to V2V communication. The lower layer characterizes the traffic flow dynamics using the Lighthill-Whitham-Richards model. The analytical solution of the asymptotic density of informed vehicles and the necessary condition for existence of the IFPW are derived for homogeneous traffic conditions. Numerical experiments provide insights on the impact of the mean communication service rate on information spread and its spatial coverage. Further, a numerical solution method is developed to solve the two-layer model, which aids in estimating the impacts of the control parameters in the queuing strategy on the IFPW speed under homogenous and heterogeneous conditions. The proposed modeling approach enables controlling the propagation of information of different information classes to meet application needs, which can assist traffic managers to design effective and efficient traffic management and control strategies under V2V communications.

## 1. Introduction

The rapid development of vehicle-to-vehicle (V2V) communication technologies has motivated their use for a wide spectrum of innovative solutions to enhance transportation safety, efficiency, and sustainability. A V2V communications-based traffic system can potentially be leveraged to enhance traffic safety by more effectively detecting emerging conflict situations, improve traffic efficiency through information-based and other control strategies, and reduce energy consumption and emissions. For example, this entails the communication of a vehicle's status to other vehicles and/or the surrounding infrastructure, and thereby the exchange of information on travel/traffic conditions. Hence, vehicles equipped with such a capability for two-way communications can potentially gain spatio-temporal knowledge on travel-related conditions, which can be used to develop vehicle-level travel strategies and/or network-level traffic management strategies. Further, V2V-based traffic systems enable decentralized information generation and dissemination. Vehicles in a V2V-based system can generate information and relay it to other vehicles through multi-hop processes. Unlike centralized information systems, a V2V-based traffic system can potentially provide timely information in emergency/disaster situations by avoiding delays associated with data collection and communication with control center.

---

[1] Corresponding author. Tel.: +1 404-894-2243.
  *Email address*: peeta@gatech.edu (S. Peeta)

Understanding the characteristics of spatiotemporal information flow propagation in a V2V-based traffic system is important as most applications require timely and reliable information delivery. However, modeling information flow propagation in space and time is challenging. Factors from both traffic flow domain and communications domain significantly affect the reliability of V2V communication and information propagation. The traffic flow dynamics affect occurrence of V2V communications. Communication constraints, such as communication frequency, channel capacity, and communication power, significantly affect the reliability of V2V communications.

In the literature, various models have been proposed to characterize information flow propagation in different traffic flow and communication environments. These models can be classified into microscopic and macroscopic. Microscopic models address information flow propagation by considering the spatial distribution of traffic flow. They factor the effects of the random positions of equipped vehicles in the traffic stream on information flow propagation. Simulation and analytical models have been developed to analyze V2V propagation performance in terms of the expected information propagation distance (Wang, 2007; Wang et al., 2010; Wang et al., 2011; Wang et al., 2012; Yin et al., 2013; Wang et al., 2015; Du and Dao, 2015; Du et al., 2016), under different traffic flow scenarios. However, these models oversimplify the effects of communication constraints by assuming that information can be retransmitted instantaneously. This assumption neglects the time latency of information flow propagation. Thereby, these models only characterize information dissemination in the space domain, but not in the time domain.

To address the aforementioned gaps, some recent studies have sought to characterize information flow propagation at a macroscopic level (Kim et al. 2017; Wang et al. 2018; Kim et al., 2018) by introducing the concept of information flow propagation wave (IFPW). They use the notion that when information spreads through multi-hop broadcasting communications, from a macroscopic perspective, there is a moving boundary that separates traffic flow into informed and uninformed regions, and moves towards the uninformed region like a wave. By leveraging the analogy of the IFPW with disease spread in epidemiology, analytical models are developed to characterize the IFPW. These macroscopic models relax the assumption of instantaneous information propagation and can describe the spatiotemporal spread of information in the traffic flow. In addition, these models capture the effects of V2V communication constraints realistically using a communication kernel. Further, interactions between V2V communications and traffic flow dynamics are captured by incorporating the effects of congested traffic, such as the backward propagating traffic wave, on information flow propagation.

The models discussed heretofore are descriptive, and seek to describe the spatiotemporal propagation of information to address effects of traffic flow dynamics and/or communication constraints. However, they lack a capability to control the propagation of information flow, which is necessary for traffic management applications in a V2V-based traffic system. For example, real-time traffic/routing information can help travelers choose better routes to reduce travel time. However, congestion can worsen if all travelers receive the same information and choose the same (or similar) routes in an uncoordinated V2V-based system or receive and choose the same routing suggestions. Hence, the propagation of information flow needs to be controlled so that the spatiotemporal access to information varies across vehicles in such a way as to improve system performance. Similarly, under emergency evacuation, the propagation speed of evacuation information needs to be controlled so that it can reach different areas in the affected region with different impact levels at the desired times to reduce the severe traffic congestion or gridlock that would otherwise occur due to the simultaneous evacuation of all evacuees.

Another common characteristic of previous studies is that they only consider the propagation of a specific information packet of interest or one type of information. In practical applications, information can belong to different classes (e.g., safety information, routing information, work zone information). Hence, a V2V-based system may need to propagate information from different information classes simultaneously. However, the application needs of information from different information classes can be different, in terms of three performance measures: (i) information spread, defined here as the proportion of vehicles informed with a specific information packet, (ii) bounds on time delays for this information to reach specific locations, and (iii) spatial coverage, defined here as the distance this information can be propagated from its point of origin. For example, urgent traffic accident information (e.g., road is blocked by an accident) needs to be delivered to all vehicles in the impacted area with low latency. By comparison, routing information needs to reach only a certain proportion of vehicles to avoid possible congestion arising from the provision of information on the suggested route. Work zone information or sudden hard brake information may need to be propagated in a small area in the vicinity of where they are generated.

This study designs a queuing-based modeling approach to control the propagation of information of different information classes to meet application needs related to information spread, time delay bounds, and spatial coverage. An information class is defined as a type of information which has similar application needs in terms of the three propagation performance measures. To enable control for multiclass information flow propagation, this study assumes that the size of each information packet is the same and the channel capacity is shared equally with all equipped vehicles within communication range of that vehicle (Wang et al., 2018). Under this assumption, an equipped vehicle can send data containing multiple information packets during one transmission, whose number is determined by the size of one information packet, channel capacity, communication frequency, and the number of equipped vehicles within communication range of that vehicle. This implies that an equipped vehicle can serve (send) multiple information packets simultaneously.

To better characterize the information service (sending) process in our queuing-based approach, we denote a "virtual communication server" (hereafter, referred to as "communication server") as the storage amount in the transmitted data that is equal to the size of an information packet. A communication server can serve at most one information packet at a time. The total number of communication servers is equal to the maximum number of information packets that an equipped vehicle can send simultaneously during one transmission, which is labeled the transmission capacity. We denote communication service time as the time duration an information packet is in the communication server. During the communication service time, the information packet will be repetitively sent by the equipped vehicle where the number of transmissions depends on the communication frequency which is the number of data transmissions per unit time enabled by the V2V device characteristics in the vehicle. We denote the mean communication service rate for a server as the inverse of the mean communication service time of all information packets served by that server.

To enable control for multiclass information flow propagation, for the first time in the literature, a queuing strategy is developed for each V2V-equipped vehicle to propagate the information packets of different information classes that it receives or generates. We assume information packets in different information classes will form different queues. Thereby, when an information packet is received by an equipped vehicle, it will be forwarded to the queue for the information class it belongs to. After being in the queue, the information packet will enter a communication server for this information class to be disseminated. It will be deleted from the server after its assigned communication service time is reached. Based on this conceptual queuing strategy, information propagation control is enabled by assigning different number of communication servers and mean communication service rates to different information classes to send the information. It should be noted that the mean communication service rate for an information class determines the mean communication service time of each information packet in the information class, which impacts the number of transmissions of each information packet in this class. Due to existence of communication failure, an information packet cannot be guaranteed to be received by other vehicles if it is just sent once by an equipped vehicle. The queuing strategy allows an equipped vehicle to control the number of transmissions of an information packet by leveraging the mean communication service rate so as to control the number of vehicles within communication range of this vehicle that can receive the information packet. Thereby, while the mean communication service rate does not impact the success rate of one V2V communication, it significantly impacts the total success rate of V2V communications by allowing an equipped vehicle to transmit information multiple times. Also, the number of communication servers assigned to an information class significantly impacts the mean waiting time in the queue for information packets in that class, which impacts the information flow propagation speed. Thereby, two control parameters, the number of assigned communication servers and mean communication service rate, can be determined for each information class to achieve the desired propagation performance related to information spread, time delay bounds, and spatial coverage.

This study conceptually extends the macroscopic models developed by Kim et al. (2017) and Wang et al. (2018), and proposes a new two-layer analytical modeling approach to characterize the IFPW under the proposed queuing strategy. An integro-differential equation (IDE) model is derived to characterize the spatiotemporal information propagation flow under the designed queuing strategy in the upper layer. The lower layer adopts the Lighthill-Whitham-Richards (LWR) model (Lighthill and Whitham, 1955; Richards, 1956) to characterize the traffic flow dynamics. The two-layer model enables investigation of the following three questions. First, what is the density of equipped vehicles that can receive a specific information packet under given values of the two control parameters? This question seeks to provide insights on controlling information spread. Second, how do the two control parameters in the queuing system impact the propagation speeds of specific information packets of interest belonging to different information classes? Addressing this question is useful for controlling the time delay of information packets of different information classes in reaching desired locations. Third, what are the conditions that can ensure the specific information packet can form a wave to be propagated over the traffic stream, and how the two control parameters impact the propagation distance of an information packet? This question addresses the necessary conditions for the formation of an IFPW which is related to the spatial coverage of information.

The contributions of this study are fivefold. First, unlike previous studies that describe how information propagates in space and time, this study newly proposes to control the spatiotemporal propagation of information to generate prescriptive solutions that can be leveraged for performance enhancement and management of V2V-based traffic systems. Second, the study addresses, for the first time, the more general case of multiple information classes that are inherent to traffic systems. To do so, it develops a queuing-based modeling approach for an equipped vehicle to propagate different types of information simultaneously. Thereby, it enables effective and efficient control for multiclass information flow propagation under different traffic and communication environments by determining the values of the two control parameters in the queuing strategy. Third, the study develops a new nonlinear IDE system to characterize the information dissemination wave. The necessary conditions for the existence of IFPW and the analytical solution for the asymptotic density of informed vehicle are derived under homogenous conditions. To the best of our knowledge, the solution of the IDE system analogy to the proposed IDE system has not been studied before, even in the epidemiology literature. These analytical expositions quantify the impacts of the two control parameters on the density of informed vehicles and the spatial coverage. Fourth, the study designs numerical solutions to solve the two-layer model under homogeneous as well as heterogeneous conditions while considering multiple performance measures. They provide valuable insights for controlling multiclass information flow propagation to achieve the desired performance in terms of information spread, the time delay to reach the target locations, and spatial coverage under heterogeneous conditions. Fifth, the study calibrates the parameters in the proposed model using NS-3 simulations, which enhances its applicability by enabling capturing the effects of communication constraints on information flow propagation more realistically.

The remainder of the paper is organized as follows. The next section discusses the designed queuing strategy and the framework of the proposed model to characterize the IFPW. Section 3 formulates a two-layer model to characterize the IFPW in space and time under the proposed queuing strategy. In Section 4, the analytically solution for the asymptotic density of informed vehicles and the condition for existence of IFPW under homogeneous traffic conditions are discussed. In addition, the numerical solution method is presented to solve the proposed two-layer model for heterogeneous conditions. Results from numerical experiments are discussed in Section 5, to demonstrate the effectiveness of the proposed model to control the propagation performance of different information classes. Section 6 provides some concluding comments.

## 2. Preliminaries

Consider a highway with a traffic flow stream consisting of V2V-equipped and V2V-unequipped vehicles. Information is generated and broadcasted to other equipped vehicles through multi-hop V2V communications. Each equipped vehicle receives information from other equipped vehicles and broadcasts such information and the information it generates to all other equipped vehicles within communication range. Let information packets relayed in the traffic flow be divided into $s$ classes, each of which has different requirements in terms of information spread, time delay bounds and spatial coverage. Let $\mathcal{L} = \{1, 2, \cdots s\}$ denotes the set of information classes. When an equipped vehicle receives multiple packets, it filters the information packets to identify those that have not been received before. It then moves such unduplicated information packets (labeled effective information packets) into the queues for the corresponding information classes to wait to be propagated according to the information class they belong to. The effective information arrival rate is affected by unsuccessful V2V communication and removal of duplicated information packets. Because such events are random and independent, following Wang et al. (2018) and Zhang et al. (2016), this study assumes that the arrival of effective information packets to different information classes follows a Poisson process. Let $\lambda_1, \lambda_2 \cdots, \lambda_s$ be the arrival rate of information packets for information classes $1, 2 \cdots, s$, respectively.

Suppose that the size of all information packets disseminated over the traffic flow is identical, and the channel capacity is shared equally with all equipped vehicles within communication range. Let $N$ denote the transmission capacity, which describes the number of information packets that can be delivered by an equipped vehicle through one V2V communication. In this study, we assume the transmission capacity for all equipped vehicles to be the same. It should be noted that $N$ has an upper bound determined by the size of an information packet, the density of the traffic flow, the communication frequency and the channel capacity (Wang et al., 2018).

To control propagation performance for each information class, a queuing strategy for relaying information of different information classes is designed in this study, as shown in Figure 1. Note that an equipped vehicle can transmit $N$ information packets during one communication. To better illustrate the queuing strategy, we assume an equipped vehicle has $N$ communication servers each of which can serve one information packet. A communication server represents the storage amount in the transmitted data that is equal to the size of one information packet (see Figure 1). The number of the communication servers assigned to a particular information class determines the maximum number of information packets in this information class that can be transmitted simultaneously by an equipped vehicle. Let $n_j$ be the number of communication servers assigned to information class $j$, $j \in \mathcal{L}$. To control multiclass information flow propagation, information packets in different classes will form different queues (see Figure 1). If one communication server for information class $j$ is empty, the first information packet in the queue for information class $j$ will enter into the server to be sent out. Let $u_j$ be the mean communication service rate (packets/second) for information packets in information class $j$. The inverse of $u_j$ (i.e., $1/u_j$) is the mean communication service time (i.e., transmission duration) for an information packet in information class $j$. The information packet in the communication server will be transmitted repetitively until the communication service time is reached. Thereby, the communication service time significantly impact the number of vehicles that can receive the specific information of interest of information class $j$.

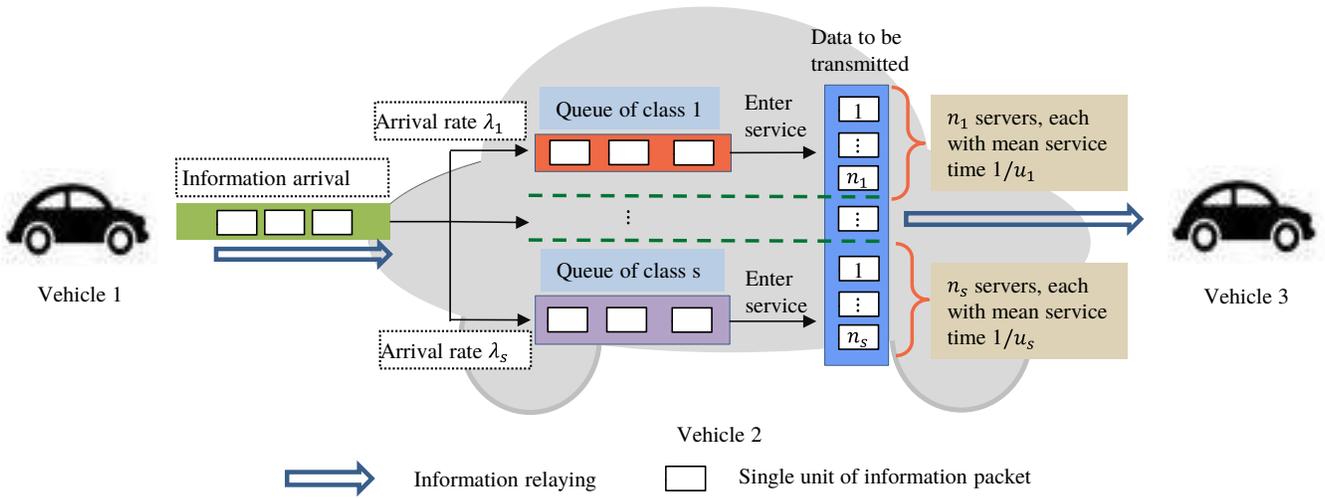

**Figure 1.** Queuing strategy for relaying information of different classes

To facilitate modeling, we assume the communication service time of each information packet in an arbitrary class $j$, $j \in \mathcal{L}$ follows an exponential distribution with mean $1/u_j$. The communication service time can be generated randomly in advance according to the exponential distribution with mean $1/u_j$. An information packet will be removed from the system if its assigned communication service time is reached. Note that for an arbitrary information class $j$, the arrival of information packets follows a Poisson process with parameter $\lambda_j$ and the corresponding communication service time follows an exponential distribution with mean $1/u_j$. Thereby, propagation of information packets in information class $j$ follows a $M/M/n_j$ queuing process.

Note that the mean communication service rate for information packets in an arbitrary information class $j$ (i.e., $u_j$) impacts the number of vehicles that can receive information packets from this information class. Further, according to the queueing theory, the mean communication service rate ($u_j$) and the number of assigned communication servers (i.e., $n_j$) for information class $j$ determine the mean waiting time of an information packet in the queue. Thereby, the propagation performance of an information packet of information class $j$ in terms of information spread, time delay bounds, and spatial coverage can be controlled by assigning various values to $n_j$ and $u_j$. It should be noted that the propagation performance of information of different information classes is constrained by the total number of communication servers in an equipped vehicle.

Under the designed queuing strategy, equipped vehicles are divided into four vehicle classes, the susceptible vehicles (labeled $S$), the information-holding vehicles (labeled $H$), the information-relay vehicles (labeled $R$) and the information-excluded vehicles (labeled $E$). Susceptible vehicles are equipped vehicles that have not received the specific information packet of interest. They become information-holding vehicles if they receive that information packet and are holding it in the queue for transmittal. The information-holding vehicles become information-relaying vehicles if that information packet enters a communication server to be disseminated to the other vehicles. Once the communication service time is reached for that information packet, it will be removed from the vehicle. The information-relying vehicle then becomes an information-excluded vehicle. It is worth noting that the susceptible vehicles can become information-relaying vehicles directly if the specific information packet of interest enters into a communication server without waiting in a queue; that is, when this information packet is received/generated, there is no queue for the corresponding information class.

Similar to Kim et al. (2017) and Wang et al. (2018), the IFPW consists of two waves: the information dissemination wave in the information flow domain and the traffic flow propagation wave in the traffic flow domain. A two-layer model is developed in this study to model the IFPW. The modeling framework is shown in Figure 2. In the upper-layer, integro-differential equations (IDEs) will be derived to characterize the information dissemination waves. This layer describes how vehicle densities by vehicle class will change instantaneously through V2V communications under the designed queuing strategy. The lower layer describes the traffic flow dynamics. In this study, the LWR model will be used to characterize traffic flow dynamics. Based on the two-layer model, the asymptotic IFPW speed, the asymptotic density of informed vehicles and the conditions for existence of IFPW will be investigated in this study.

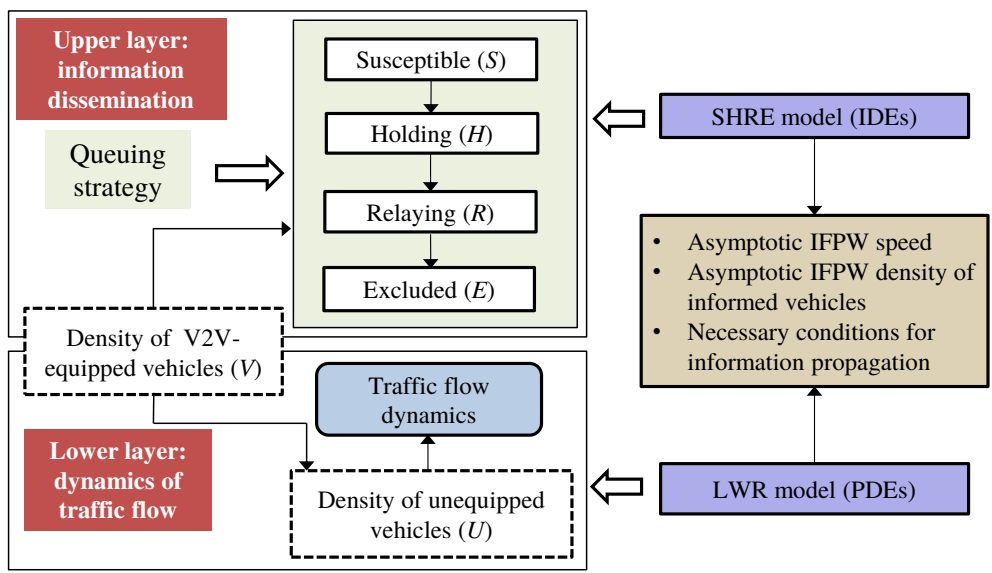

Figure 2. IFPW modeling framework under the designed queuing strategy

It should be noted that while the starting point for the modeling framework in this study is similar to those of Kim et al. (2017) and Wang et al. (2018), it differs from them fundamentally in four key aspects. First, while Kim et al. (2017) and Wang et al. (2018) only descriptively quantify the impacts of traffic flow and communication constraints on information flow propagation, this study seeks to control information flow propagation to ensure that performance in terms of information spread, time delay bounds and spatial converge can meet the application needs of information under different traffic flow and communication environments. Second, this study proposes a queuing strategy for multiclass information flow propagation that enables distribution of limited communication resources in an equipped vehicle (i.e., number of communication servers, and mean communication service time) for sending information of different classes. This improves information propagation efficiency by coordinating the communication resources with dynamic information arrival rates. However, the communication service time, which is critical for V2V communication reliability, is not controlled in Kim et al. (2017) and Wang et al. (2018). For example, the communication service time may be too short when information arrival rate is high, precluding reliable propagation of that information. Third, this study develops a new nonlinear IDE system to characterize the information flow dissemination wave in the upper layer, and analytically derives its solution. This IDE system has not been studied before even in epidemiology, unlike the IDE systems developed in Kim et al. (2017) and Wang et al. (2018) which have been extensively studied in epidemiology. Fourth, unlike those studies, this study verifies the success rate of V2V communications and calibrates parameters using NS-3 simulations. This enhances the applicability of the proposed model by better capturing the impacts of traffic flow and communication constraints on V2V communications.

## 3. Modeling the multiclass information flow propagation wave

### 3.1 Modeling the information flow dissemination wave in the upper layer

Assume the specific information packet of interest belongs to an arbitrary information class $j$, $j \in \mathcal{L}$. This section seeks to model the information dissemination wave in the upper layer under the designed queuing strategy. It describes the instantaneous change in the density of equipped vehicles by vehicle class (i.e., $S, H, R, E$ for information packets in class $j$) due to V2V communications. The impacts of communication constraints (communication power, communication frequency, signal interference, etc.) on the success of V2V communications is explicitly factored in this model.

Let $t$ be the current time. Divide the time horizon of interest uniformly into consecutive time windows of length $w$ each. Denote $S_j(x,t)$ and $I_j(x,t)$ as the densities of the susceptible vehicles and the vehicles informed with the specific information packet of interest in information class $j$, respectively, at time $t$ and location $x$. Note that the informed vehicles consist of vehicles from classes $H, R$ and $E$ which have received this information packet. Let $\Delta S_j(x,t) = S_j(x,t) - S_j(x,t-w)$ be the density of vehicles informed with the specific information packet of interest during time $[t-w,t]$. Denote $\Delta I_j(y, t-w) = I_j(x,t) - I_j(x, t-w)$ as the density change of informed vehicles at location $x$ during time interval $[t-w,t]$. Note $\Delta I_j(y, t-w) = -\Delta S_j(x, t-w)$ as an equipped vehicle is either susceptible or informed. Let the current time be $t = m \cdot w$; $m$ is a positive integer. Conceptually adapting from epidemiology and modifying, $\Delta S_j(x,t)$ is formulated as:

$$\Delta S_j(x,t) = -S_j(x, t-w) \cdot \int_\Omega K(x,y) \cdot w\beta \cdot R_j(y,t) dy \tag{1}$$

where $\Omega$ denotes the domain of space, $\beta$ is communication frequency, and $w\beta$ denotes the expected number of transmissions occurring in time length $w$. Also, $R_j(y,t)$ is the density of vehicles relaying the specific information packet of interest in information class $j$ at time $t$ and location $y$. Function $K(x,y)$ is a communication kernel which represents the probability that a susceptible vehicle at location $y$ can successfully receive the specific information packet sent from a vehicle at location $x$ under a given communication environment (communication frequency, channel capacity, communication power, etc.) and traffic flow environment (traffic density, etc.). It characterizes the reliability of V2V communication realistically by capturing the impact of factors in both the communication and traffic flow domains. In the study experiments in Section 5, function $K(x,y)$ is calibrated using NS-3 simulation. $\int_\Omega K(x,y) \cdot w\beta \cdot R_j(y,t) dy$ denotes the probability that a susceptible vehicle at location $x$ receives the specific information packet of information class $j$ sent by an informed vehicle over the space domain $\Omega$.

Suppose the mean arrival rate of information packets in information class $j$, $j \in \mathcal{L}$ is $\lambda_j$ packets/second. Let $n_j$ and $u_j$ be the number of communication servers and the mean communication service rate (packets/second) assigned to information class $j$, respectively. $n_j$ and $u_j$ are controllable parameters which will be leveraged to control the propagation of the specific information packet of interest of information class $j$. To ensure the information packets in information class $j$ can be propagated by an equipped vehicle, $n_j$ and $u_j$ are selected such that $\lambda_j < n_j u_j$, $\forall j \in \mathcal{L}$.

As discussed earlier, under the designed queuing strategy, the arrival and service process of information packets in information class $j$ follow $M/M/n_j$ queue process. Thereby, at current time $t$, the vehicles that are relaying the specific information packet of interest of information class $j$ consist of two groups: (1) the vehicles relaying the specific information packet without the information-holding process (i.e., queuing process). This implies when these vehicles receive the specific information packet, there is no queue for information class $j$ in these vehicles. Thereby, the specific information packet of interest can enter into the communication server to be disseminated out directly; and (2) the vehicles relaying the specific information packet with an information-holding process (queue process), i.e., the specific information packet of interest experiences a queuing process before entering into the communication server.

Let $T_j^q$ be the waiting time for the specific information packet of interest of information class $j$, and $W_j^q(v_j) = \Pr\{T_j^q \leq v\}$ be the probability that the waiting time of this information packet in the queue is less than $v$. According to Gross et al. (2008, page 71), we have

$$W_j^q(v) = \Pr\{T_j^q \leq v\} = 1 - \frac{r_j^{n_j} P_j^0}{n_j!(1-\rho_j)} e^{-(n_j u_j - \lambda_j) v_j} \tag{2}$$

where $r_j = \lambda_j/u_j$; $\rho_j = \lambda_j/(n_j u_j)$; and $P_j^0$ is the probability that there is no information packet of information class $j$ in the system, formulated as

$$P_j^0 = \left( \frac{r_j^{n_j}}{n_j!(1-\rho_j)} + \sum_{l=0}^{n_j-1} \frac{r_j^l}{l!} \right)^{-1} \tag{3}$$

According to Eq. (2),

$$W_j^q(0) = \Pr\{T_j^q \leq 0\} = \Pr\{T_j^q = 0\} = 1 - \frac{r_j^{n_j} P_j^0}{n_j!(1-\rho_j)} \tag{4}$$

where $W_j^0$ is the probability that the specific information packet is received by a vehicle at a time instant when there is no queue for information class $j$. Let

$$\xi_j = \frac{r_j^{n_j} P_0}{n_j!(1-\rho_j)} \tag{5}$$

Then

$$W_j^q(0) = 1 - \xi_j \tag{6}$$

From Eq. (2), we have

$$\Pr\{T_j^q > v\} = 1 - W_j^q(v) = \frac{r_j^{n_j} P_0}{n_j!(1-\rho_j)} e^{-(n_j u_j - \lambda_j) v_j} \tag{7}$$

Let $T_j^s$ be the service time of the specific information packet of interest in the communication server, and $\Pr\{T_j^q \leq \theta\}$ be the probability that the communication service time is less than $\theta$. Recall the communication service time follows exponential distribution with mean $1/u_j$, then

$$\Pr\{T_j^q \leq \theta\} = 1 - e^{-u_j\theta} \tag{8}$$

This implies

$$\Pr\{T_j^q > \theta\} = e^{-u_j\theta} \tag{9}$$

$R_j(y,t)$ can then be formulated as

$$R_j(y,t) = A_j(y,t) + B_j(y,t) \tag{10}$$

where

$$A_j(y,t) = W_j^q(0) \cdot \sum_{i=1}^{t/w} \Delta I_j(y, t-iw) \cdot \Pr\{T_j^s > iw\}$$

$$= (1-\xi_j) \cdot \sum_{i=1}^{t/w} \Delta I_j(y, t-iw) \cdot e^{-u_j \cdot iw}$$

$$B_j(y,t) = \sum_{i=1}^{t/w} \int_0^{iw} \Delta I_j(y, t-iw) \cdot \frac{\partial\left(\Pr\{T_j^q > v\}\right)}{\partial v} \cdot \Pr\{T_j^s > iw - v\} dv$$

$$= \sum_{i=1}^{t/w} \int_0^{iw} \Delta I_j(y, t-iw) \cdot (n_j u_j - \lambda_j) \cdot \xi_j \cdot e^{-(n_j u_j - \lambda_j)v} \cdot e^{-u_j \cdot (iw - v)} dv$$

where $A_j(y,t)$ is the accumulated density of vehicles relaying the specific information packet of interest of information class $j$ without queuing process at location $y$ and current time $t$. $\Delta I_j(y, t-iw) \cdot \Pr\{T_j^s > iw\}$ is the density of vehicles informed at time $t - iw$ and relaying the specific information packet at location $y$ and current time $t$. $B_j(y,t)$ is the accumulated density of vehicles relaying the specific information packet after queuing process at location $y$ and current time $t$. The term $\int_0^{iw} \Delta I_j(y, t-iw) \cdot \frac{\partial\left(\Pr\{T_j^q > v\}\right)}{\partial v} \cdot \Pr\{T_j^s > iw - v_j\} dv$ denotes the density of vehicles at location $y$ that become informed $iw$ time units ago and are propagating the specific information at current time $t$ after experiencing the queuing process.

To derive the continuous model, let $w \to 0$; dividing both sides of Eq. (1) by $w$, we have

$$\lim_{w \to 0} \frac{\Delta S_j(x,t)}{w} = \frac{\partial S_j(x,t)}{\partial t} = \lim_{w \to 0} -\beta S_j(x, t-w) \cdot \int_\Omega K(x,y) \cdot w \frac{(A_j(y,t) + B_j(y,t))}{w} dy \tag{11}$$

Note

$$\Delta I_j(y, t-iw) = I_j(y, t-(i-1)w) - I_j(y, t-iw) \approx \frac{\partial I_j(y, t-iw)}{\partial t} w. \tag{12}$$

Then

$$\lim_{w \to 0} -\beta S_j(x, t-w) \cdot \int_\Omega K(x,y) \cdot w \cdot A_j(y,t) dy$$

$$= \lim_{w \to 0} -\beta(1-\xi_j) S_j(x, t-w) \cdot \sum_{i=1}^{t/w} \int_\Omega \frac{\partial I_j(y, t-iw)}{\partial t} \cdot K(x,y) \cdot w \cdot e^{-u_j \cdot iw} dy. \tag{13a}$$

and

$$\lim_{w \to 0} -\beta S_j(x, t-w) \cdot \int_\Omega K(x,y) \cdot w \cdot B_j(y,t) dy$$
$$= \lim_{w \to 0} -\beta S_j(x, t-w) \cdot \int_\Omega K(x,y) \cdot f_j(y, iw) dy \tag{13b}$$

where

$$f_j(y, iw) = \sum_{i=1}^{t/w} \int_0^{iw} \frac{\partial I_j(y, t-iw)}{\partial t} w \cdot (n_j u_j - \lambda_j) \xi_j \cdot e^{-(n_j u_j - \lambda_j)v} \cdot e^{-u_j \cdot (iw - v)} dv \tag{13c}$$

Note the terms $(\partial I_j(y, t-\tau)/\partial t) K(x,y) e^{-\lambda \tau}$ and $\int_0^\tau \frac{\partial I(y,t-\tau)}{\partial t} \cdot (n_j u_j - \lambda_j) \cdot \xi_j \cdot e^{-(n_j u_j - \lambda_j)v} \cdot e^{-u_j \cdot (\tau-v)} dv$ are continuous and bounded in the time domain. Both of them are Riemann integrable. Thereby, Eq. (13a) and Eq. (13b) can be written, respectively, as:

$$\lim_{w \to 0} -\beta(1-\xi_j) S_j(x, t-w) \cdot \sum_{i=1}^{t/w} \int_\Omega \frac{\partial I_j(y, t-iw)}{\partial t} \cdot K(x,y) \cdot w \cdot e^{-u_j iw} dy$$
$$= -\beta \cdot (1-\xi_j) \cdot S_j(x,t) \cdot \int_\Omega \int_0^t \frac{\partial I_j(y, t-\tau)}{\partial t} \cdot K(x,y) \cdot e^{-u_j \tau} d\tau \cdot dy \tag{14a}$$

$$\lim_{w \to 0} -\beta S_j(x, t-w) \cdot \int_\Omega K(x,y) \cdot f_j(y, iw) dy$$
$$= \beta S_j(x,t) \cdot \int_\Omega \int_0^t \int_0^\tau \frac{\partial I_j(y, t-\tau)}{\partial t} K(x,y) (n_j u_j - \lambda_j) \xi_j \cdot e^{-(n_j u_j - \lambda_j)v} e^{-u_j \cdot (\tau-v)} dv \cdot d\tau \cdot dy \tag{14b}$$

Note $\partial I_j(x, t-\tau)/\partial t = -\partial S_j(x, t-\tau)/\partial t$; substituting Eq. (14) into Eq. (11), we have

$$\frac{\partial S_j(x,t)}{\partial t} = \beta \cdot (1-\xi_j) \cdot S_j(x,t) \cdot \int_\Omega \int_0^t \frac{\partial S_j(y, t-\tau)}{\partial t} \cdot K(x,y) \cdot e^{-u_j \tau} d\tau dy +$$
$$\beta S_j(x,t) \cdot \int_\Omega \int_0^t \int_0^\tau \frac{\partial I_j(y, t-\tau)}{\partial t} K(x,y) (n_j u_j - \lambda_j) \xi_j \cdot e^{-(n_j u_j - \lambda_j)v} e^{-u_j \cdot (\tau-v)} dv \cdot d\tau \cdot dy \tag{15}$$

According to Eq. (10) and Eq. (12), in continuous space, the density of information-relaying vehicles can be written as

$$R_j(y,t) = -(1-\xi_j) \int_0^t \frac{\partial S_j(y, t-\tau)}{\partial t} \cdot e^{-u_j \tau} d\tau -$$
$$\int_0^t \int_0^\tau \frac{\partial S_j(y, t-\tau)}{\partial t} \cdot (n_j u_j - \lambda_j) \xi_j \cdot e^{-(n_j u_j - \lambda_j)v} \cdot e^{-u_j \cdot (\tau-v)} dv \cdot d\tau \tag{16}$$

The terms $-(1-\xi_j) \int_0^t \frac{\partial S_j(y,t-\tau)}{\partial t} \cdot e^{-u_j \tau} d\tau dy$ and $-\int_0^t \int_0^\tau \frac{\partial S_j(y,t-\tau)}{\partial t} \cdot (n_j u_j - \lambda_j) \xi_j \cdot e^{-(n_j u_j - \lambda_j)v} \cdot e^{-u_j \cdot (\tau-v)} dv \cdot d\tau$ denote the density of information-relaying vehicles without queuing process and with queuing process, respectively. Let $\eta = t - \tau$, then

$$R_j(y,t) = f_{j,1}(y,t) + f_{j,2}(y,t) \tag{17a}$$

where

$$f_{j,1}(y,t) = -(1-\xi_j) \int_0^t \frac{\partial S_j(y,\eta)}{\partial \eta} \cdot e^{-u_j(t-\eta)} d\eta \tag{17b}$$

$$f_{j,2}(y,t) = -\int_0^t \int_0^{t-\eta} \frac{\partial S_j(y,\eta)}{\partial \eta} \cdot (n_j u_j - \lambda_j)\xi_j \cdot e^{-(n_j u_j - \lambda_j)v} \cdot e^{-u_j \cdot (t-\eta-v)} dv \cdot d\eta \quad (17c)$$

The derivative of $R_j(y,t)$ with respect to $t$ is

$$\frac{\partial R_j(y,t)}{\partial t} = \frac{\partial f_{j,1}(y,t)}{\partial t} + \frac{\partial f_{j,2}(y,t)}{\partial t} \quad (18a)$$

where

$$\frac{\partial f_{j,1}(y,t)}{\partial t} = -(1-\xi_j)\frac{\partial S_j(y,t)}{\partial t} + u_j(1-\xi_j)\int_0^t \frac{\partial S_j(y,\eta)}{\partial \eta} \cdot e^{-u_j(t-\eta)} d\eta$$

$$= -(1-\xi_j)\frac{\partial S_j(y,\eta)}{\partial t} - u_j f_{j,1}(y,t) - u_j R_j(y,t) \quad (18b)$$

$$\frac{\partial f_{j,2}(y,t)}{\partial t} = -\int_0^t \frac{\partial S_j(y,\eta)}{\partial \eta} \cdot (n_j u_j - \lambda_j)\xi_j e^{-(n_j u_j - \lambda_j)(t-\eta)} d\eta + u_j f_{j,1}(y,t) \quad (18c)$$

Thereby,

$$\frac{\partial R_j(y,t)}{\partial t} = -(1-\xi_j)\frac{\partial S_j(y,\eta)}{\partial t} - \int_0^t \frac{\partial S_j(y,\eta)}{\partial \eta} \cdot (n_j u_j - \lambda_j)\xi_j \cdot e^{-(n_j u_j - \lambda_j)(t-\eta)} d\eta \quad (19)$$
$$- u_j R_j(y,t)$$

Let $H_j(y,t) = -\int_0^t \frac{\partial S_j(y,\eta)}{\partial \eta} \cdot \xi_j \cdot e^{-(n_j u_j - \lambda_j)(t-\eta)} d\eta$. According to Eq. (7), the probability that an information packet in information class $j$ is received by a vehicle at time $\eta$ and is waiting in the queue at current time $t$ $(t > \eta)$ is $\xi_j \cdot e^{-(n_j u_j - \lambda_j)(t-\eta)}$. This implies that $H_j(y,t)$ is the density of information-holding vehicles at location $y$ and time $t$. Differentiating $H_j(y,t)$ with respect to $t$, we have

$$\frac{\partial H_j(y,t)}{\partial t} = -\xi_j \frac{\partial S_j(y,t)}{\partial t} - (n_j u_j - \lambda_j) H_j(y,t) \quad (20)$$

Let $E_j(x,t)$ denote the density of information-excluded vehicles at location $x$ at time $t$. As informed vehicles consist of the information-holding, information-relaying and information-excluded vehicles, $I_j(x,t) = H_j(x,t) + R_j(x,t) + E_j(x,t)$. Thereby, we have

$$\frac{\partial H_j(x,t)}{\partial t} + \frac{\partial R_j(x,t)}{\partial t} + \frac{\partial E_j(x,t)}{\partial t} = \frac{\partial I_j(x,t)}{\partial t} = -\frac{\partial S_j(x,t)}{\partial t}. \quad (21)$$

Substituting Eq. (19) and Eq. (20) into Eq. (21), yields

$$\frac{\partial E_j(x,t)}{\partial t} = u_j R_j(x,t) \quad (22)$$

According to the above analysis, we have the following IDE system:

$$\begin{cases} \frac{\partial S_j(x,t)}{\partial t} = -\beta S_j(x,t) \int_\Omega R_j(y,t) \cdot K(x,y) dy & (23a) \\ \frac{\partial H_j(x,t)}{\partial t} = \beta \cdot \xi_j \cdot S_j(x,t) \int_\Omega R_j(y,t) \cdot K(x,y) dy - (n_j u_j - \lambda_j) \cdot H_j(x,t) & (23b) \\ \frac{\partial R_j(x,t)}{\partial t} = (1-\xi_j) \cdot \beta S_j(x,t) \int_\Omega R_j(y,t) \cdot K(x,y) dy + (n_j u_j - \lambda_j) \cdot H_j(y,t) - u_j \cdot R_j(x,t) & (23c) \\ \frac{\partial E_j(x,t)}{\partial t} = u_j \cdot R_j(x,t) & (23d) \end{cases}$$

For simplicity, we will label the IDE system (23) as the susceptible-holding-relaying-excluded (SHRE) model. It describes the instantaneous change in densities of vehicles by vehicle class for dissemination of an information packet in information class $j$. Eq. (23) shows that susceptible vehicles become informed vehicles at a rate proportional to the densities of susceptible vehicles and information-relaying vehicles (see Eq. (23a)). According to Eq. (23b), information-holding vehicles become information-relaying vehicles at a rate inversely proportional to $(n_j u_j - \lambda_j)$. Thereby, if the assigned number of communication servers ($n_j$) and the mean communication service rate ($u_j$) are increased, information-holding vehicles would become information-relaying vehicles faster. This implies that the specific information packet of interest experiences less waiting time in the queue. Hence, it can be propagated in the traffic stream at a higher speed. Eq. (23c) indicates that the density change of information-relaying vehicle increases monotonically with respect to number of communication servers ($n_j$) and the mean communication service rate ($u_j$). According to Eq. (23d), the density change of information-excluded vehicles is proportional to $u_j$. Note $1/u_j$ denotes the mean communication service time of information packets of information class $j$. If $u_j$ is smaller, the information packet would stay in the commination server for a longer time, implying that it can be disseminated more times using the repetitive broadcast process. This will impact both the IFPW speed and asymptotic density of informed vehicles. Thereby, we can control $n_j$ and $u_j$ to meet the application needs of information packets in information class $j$.

It is worth noting that Eq. (23d) can be used to characterize the dissemination wave of information packets in an arbitrary information class $j$, $j \in \mathcal{L}$. As $\sum_{i=1}^{S} n_i = N$, we can assign different number of communication servers and mean communication service rates for different information classes appropriately to meet their application needs simultaneously. Eq. (23) also implies that if $\xi_j \equiv 0$ (i.e., $P_j^0 = 0$, or $n_j u_j \to \lambda_j$), the SHRE model becomes the susceptible-relaying-excluded model studied by Wang et al. (2018). It models the information flow dissemination wave under an information-relay control strategy where there is no queuing delay (i.e., no information-holding vehicles).

**3.2 Modeling the traffic flow dynamics in the lower layer**

The upper-layer SHRE model describes how the density of vehicles by vehicle class changes instantaneously due to V2V communications. It captures the impacts of communication constraints on success rate of V2V communications, as also factors the effects of the queuing strategy and the distribution of information-relaying vehicles on the IFPW formation. As mentioned before, the IFPW is a combination of the information flow dissemination wave and the traffic flow propagation wave. This section models the traffic flow dynamics to determine the traffic flow propagation wave. The effects of traffic flow dynamics on IFPW are threefold. First, they impact the success rate of V2V communications (i.e., $K(x, y)$) by determining the number of equipped vehicles within communication range of an information-relaying vehicle. Second, the number of the vehicles sending the specific information depends on the spatial distribution of information-relaying vehicles which evolves with the traffic flow dynamics. Third, the traffic flow speed significantly contributes to the IFPW speed. It adds to the IFPW speed in the direction of vehicular traversal and reduces the IFPW speed in direction opposite to that of vehicular traversal.

In this study, the first-order LWR model is used to describe the traffic flow dynamics. It can reproduce some essential features of traffic flow, such as the formation and propagation of traffic flow waves. The model consists of the flow conservation law and an explicit density-flow relationship known as the fundamental diagram of traffic flow. The flow conservation law and the fundamental diagram can be expressed as the following PDE model:

$$\frac{\partial k(x,t)}{\partial t} + \frac{\partial q(x,t)}{\partial x} = 0 \tag{24}$$

$$q(x,t) = F(k,x,t) \tag{25}$$

where $k(x,t)$ is the traffic flow density at location $x$ at time $t$, $q(x,t)$ is the instantaneous flow, and $F(k,x,t)$ is the fundamental diagram in which the flow and density are related by a continuous and piecewise differentiable equation.

## 4. Analytical and numerical solutions for the two-layer model

This section discusses the analytical and numerical solutions for the two-layer model under homogeneous and heterogeneous traffic conditions, respectively. Homogeneous traffic conditions refer to unidirectional traffic flow with uniform traffic flow density and the heterogeneous traffic conditions refer to the other traffic flow situations (e.g., bi-directional flow). Under homogeneous traffic conditions, the traffic flow dynamics only shift the IFPW towards the direction of traffic flow and do not change the densities of vehicles of different classes. Thereby, the impacts of traffic flow dynamics on the IFPW speed are uniform in space and time. The asymptotic density of informed vehicles and the condition for existence of IFPW can be derived analytically using only the upper-layer SHRE model.

Under heterogeneous traffic flow conditions, the traffic flow dynamics change the densities of vehicles of different classes spatiotemporally. Thereby, the impacts of the traffic flow dynamics on the IFPW speed are non-uniform in space and time. To obtain the solutions of the two-layer model under heterogeneous conditions, the change in density of vehicles of each vehicle class due to V2V communications (in the upper layer) and the traffic flow dynamics (in the lower layer) need to be tracked simultaneously. To do so, a numerical solution method is proposed here to capture the interactions between the upper and lower layers sequentially under discrete time and space settings.

### 4.1 Analytical solution of the two-layer model under homogeneous traffic conditions

The proposed SHRE model conceptually adapts the idea of susceptible-exposed-infected-recovered (SEIR) model that is extensively studied in epidemiology (see McCluskey, 2012; Li and Muldowney, 1995; Li et al., 1999; Smith et al., 2001). The equilibrium solution and conditions for local stability of the SEIR model are analyzed in these studies. However, the classical SEIR only address the temporal spread of a disease among the population at one location. By comparison, the designed SHRE model is a spatial model that seeks to determine how a specific information packet of interest will be propagated in both space and time. Thereby, the solutions of SEIR model in previous epidemiological studies cannot be applied to the SHRE model. To the best of our knowledge, the solution of the IDE system analogy to the SHRE model has not been studied before. Here, we derive analytical solution for the asymptotic density of vehicles by vehicle class of the SHRE model and study the conditions for the existence of the IFPW.

Let $\sigma$ be the density of equipped vehicles. Suppose at time 0, all equipped vehicles are susceptible vehicles in the highway. Thereby, the initial conditions for the SHRE model are $S_j(x,0) = \sigma$, $H_j(x,0) = R_j(x,0) = E_j(x,0) = 0$. Assume the specific information packet of interest of information class $j$ is generated and propagated by an equipped vehicle at location 0 and time 0. Similar to Kim et al. (2017) and Wang et al., (2017), under homogeneous traffic conditions, the information under the designed queuing strategy will quickly form a wave (if it exists) to propagate backward and forward at a uniform speed. The asymptotic density of vehicles of each vehicle class is the same beyond the location where the wave speed is stable. However, we cannot derive the analytical solutions for asymptotic speed of the IFPW. The IFPW speed will be solved for using the numerical method introduced in Section 4.2.

As discussed in Section 3.1, the communication kernel $K(x,y)$ characterizes the one-hop success rate of V2V communications. It significantly impacts the traveling wave solutions. In this study, the Gaussian communication kernel in Eq. (26) proposed by Kim et al. (2017) will be used to characterize the one-hop success rate of V2V communications, in which the parameters $a$ and $b$ are calibrated using V2V communication data obtained through NS-3 simulation. It shows that the communication is subject to attenuation over distance. Also, it can be noted that the communication kernel satisfies $\int_\Omega K(x,y)dy = \int_{-\infty}^{+\infty} K(x,y)dy = b$.

$$K(x,y) = \frac{b}{a\sqrt{\pi}} e^{\frac{-(x-y)^2}{a^2}}, a > 0, 0 < b \leq 1, \tag{26}$$

The asymptotic density of vehicles by vehicle class and asymptotic density of informed vehicles are defined as follows.

**Definition 1** (asymptotic density of vehicles by vehicle class): the asymptotic density of vehicles of vehicle class $z_j^*(x)$, $z_j \in \{S_j, H_j, R_j, E_j\}$ at location $x$ is defined as $z_j^*(x) = \lim_{t \to \infty} z(x,t)$.

**Definition 2** (asymptotic density of informed vehicles): let $I_j^*(x)$ be the asymptotic density of informed vehicles at location $x$, it is defined as $I_j^*(x) = \lim_{t \to \infty} I_j(x,t)$.

The following two theorems will be useful to analyze the asymptotic density of informed vehicles.

**Theorem 1** (asymptotic density of information-holding vehicles): if $n_j u_j > \lambda_j$, then $H_j^*(x) = \lim_{t \to +\infty} H_j(x,t) = 0$

*Proof*: Let $\omega_j = n_j u_j - \lambda_j$; multiplying both sides of Eq. (23b) by $e^{\omega_j t}$, we have

$$e^{\omega_j t} \frac{\partial H_j(x,t)}{\partial t} + e^{\omega_j t} \cdot H_j(x,t) = e^{\omega_j t} \beta \cdot \xi_j \cdot S_j(x,t) \int_\Omega R_j(y,t) \cdot K(x,y) dy \tag{27}$$

This implies

$$\frac{\partial \left( e^{\omega_j t} H_j(x,t) \right)}{\partial t} = e^{\omega_j t} \beta \cdot \xi_j \cdot S_j(x,t) \int_\Omega R_j(y,t) \cdot K(x,y) dy \tag{28}$$

Then

$$e^{\omega_j t} H_j(x,t) - e^{\omega_j t} H_j(x,0) = \int_0^t e^{\omega_j \tau} \beta \cdot \xi_j \cdot S_j(x,\tau) \int_\Omega R_j(y,\tau) \cdot K(x,y) dy \cdot d\tau \tag{29}$$

Note $H_j(x,0) \equiv 0$. According to Eq. (23d), Eq. (29) can be written as

$$\begin{aligned} H_j(x,t) &= e^{-\omega_j t} \beta \cdot \xi_j \cdot \int_0^t e^{\omega_j \tau} S_j(x,\tau) \int_\Omega \frac{1}{u_j} \frac{\partial E_j(y,\tau)}{\partial \tau} \cdot K(x,y) dy \cdot d\tau \\ &= \frac{e^{-\omega_j t} \beta \cdot \xi_j}{u_j} \int_\Omega K(x,y) \int_0^t S_j(x,\tau) e^{\omega_j \tau} \frac{\partial E_j(y,\tau)}{\partial \tau} d\tau \cdot dy \end{aligned} \tag{30}$$

Note that as $t \to +\infty$, the densities of vehicles of different vehicle classes at each location $y$ become stable. Thereby, $\lim_{t \to +\infty} \partial E_j(y,\tau)/\partial t \to 0$. For an arbitrarily small positive value $\varepsilon$, let $g(g < +\infty)$ be the value such that $\partial E_j(y,\tau)/\partial \tau < \varepsilon$ for $\tau > g$. Then

$$\lim_{t \to +\infty} H_j(x,t) = \lim_{t \to +\infty} \frac{e^{-\omega_j t}\beta \cdot \xi_j}{u_j} \int_\Omega K(x,y) \int_0^t S_j(x,\tau) e^{\omega_j \tau} \frac{\partial E_j(y,\tau)}{\partial \tau} d\tau \cdot dy$$

$$= \lim_{t \to +\infty} \frac{e^{-\omega_j t}\beta \cdot \xi_j}{u_j} \int_\Omega K(x,y) \int_0^g S_j(x,\tau) e^{\omega_j \tau} \frac{\partial E_j(y,\tau)}{\partial \tau} d\tau \cdot dy \quad (31)$$

$$+ \lim_{t \to +\infty} \frac{e^{-\omega_j t}\beta \cdot \xi_j}{u_j} \int_\Omega K(x,y) \int_g^{+\infty} S_j(x,\tau) e^{\omega_j \tau} \frac{\partial E_j(y,\tau)}{\partial \tau} d\tau \cdot dy$$

Note that $S_j(x,\tau) e^{\omega_j \tau} \frac{\partial E_j(y,\tau)}{\partial \tau}$ is bounded when $\tau \in [0,g]$. Thereby, $\int_0^g S_j(x,\tau) e^{\omega_j \tau} \frac{\partial E_j(y,\tau)}{\partial \tau} d\tau$ is bounded. This implies

$$\lim_{t \to +\infty} \frac{e^{-\omega_j t}\beta \cdot \xi_j}{u_j} \int_\Omega K(x,y) \int_0^g S_j(x,\tau) e^{\omega_j \tau} \frac{\partial E_j(y,\tau)}{\partial \tau} d\tau \cdot dy \to 0 \quad (32)$$

As $\partial E_j(y,\tau)/\partial \tau < \varepsilon$ for $\tau > g$. Then

$$\lim_{t \to +\infty} \frac{e^{-\omega_j t}\beta \cdot \xi_j}{u_j} \int_\Omega K(x,y) \int_g^{+\infty} S_j(x,\tau) e^{\omega_j \tau} \frac{\partial E_j(y,\tau)}{\partial \tau} d\tau \cdot dy$$

$$\leq \lim_{t \to +\infty} \frac{e^{-\omega_j t}\beta \cdot \xi_j}{u_j} \sigma\, e^{\omega_j t} \cdot \varepsilon \quad (33)$$

$$= \frac{\beta \cdot \xi_j}{u_j} \sigma \cdot \varepsilon$$

Note that $\varepsilon$ is an arbitrarily small positive value. Thereby,

$$H_j^*(x) = \lim_{t \to +\infty} H_j(x,t) = 0 \quad (34)$$

**Theorem 2** (asymptotic density of information-relaying vehicles): if $n_j u_j > \lambda_j$, then $R_j^*(x) = \lim_{t \to +\infty} R_j(x,t) = 0$.

Theorem 2 can be proved using the same method used to prove Theorem 1; it is omitted here to avoid duplication.

Theorems 1 and 2 indicate that if $n_j u_j > \lambda_j$, there would be no information-holding and information-relaying vehicles at each location eventually. This is because when $n_j u_j > \lambda_j$, the specific information packet of interest waiting in the queue for information class $j$ will have finite waiting time and finite communication service time. It will enter into the communication server for propagation and is removed from it eventually.

Let $\gamma_j = \beta b \sigma/u_j$, the following theorem discusses the asymptotic density of information-excluded vehicles.

**Theorem 3** (asymptotic density of information-excluded vehicles): if $n_j u_j > \lambda_j$, and $\gamma_j > 1$, then $R_j^*(x) = \lim_{t \to +\infty} R_j(x,t) = \sigma \cdot \alpha_j^*$, where $\alpha_j^* \in (0,1)$ is the unique solution of the following nonlinear equation

$$e^{-\gamma_j \alpha_j} + \alpha_j - 1 = 0 \quad (35)$$

*Proof*: Under homogeneous traffic conditions, the traffic flow density is uniform. Hence, for arbitrary time $t$ and location $x$, we have

$$S_j(x,t) + H_j(x,t) + R_j(x,t) + E_j(x,t) = \sigma \quad (36)$$

Note that $E_j(x,0) = 0$; according to Eq. (23d):

$$E_j(x,t) = \int_0^t u_j \cdot R_j(y,t) dt \tag{37}$$

Eq. (23a) implies that

$$\frac{\partial \left( \ln\left(S_j(x,t)\right) \right)}{\partial t} = -\beta \int_\Omega R_j(y,t) \cdot K(x,y) dy \tag{38}$$

Integrating both sides of Eq. (38) from 0 to $t$, we have

$$\ln\left(S_j(x,t)\right) - \ln(\sigma) = -\beta \int_\Omega \left[ \int_0^t R_j(y,\tau) d\tau \right] \cdot K(x,y) dy \tag{39}$$

Substituting Eq. (37) into Eq. (39), yields

$$\ln\left(S_j(x,t)\right) - \ln(\sigma) = -\frac{\beta}{u_j} \int_\Omega \left[ \int_0^t u_j R_j(y,\tau) d\tau \right] \cdot K(x,y) dy$$

$$= -\frac{\beta}{u_j} \int_\Omega E_j(x,t) \cdot K(x,y) dy \tag{40}$$

Thereby,

$$S_j(x,t) = \sigma \cdot e^{-\frac{\beta}{u_j} \int_\Omega E_j(x,t) \cdot K(x,y) dy} \tag{41}$$

Substituting Eq. (41) into Eq. (36), we have

$$\sigma \cdot e^{-\frac{\beta}{u_j} \int_\Omega E_j(x,t) \cdot K(x,y) dy} + H_j(x,t) + R_j(x,t) + E_j(x,t) = \sigma \tag{42}$$

Let $t \to +\infty$, then

$$\sigma \cdot e^{-\frac{\beta}{u_j} \int_\Omega E_j^*(x) \cdot K(x,y) dy} + H_j^*(x) + R_j^*(x) + E_j^*(x) = \sigma \tag{43}$$

Note that $\int_\Omega E_j^*(x) \cdot K(x,y) dy = b \cdot E_j^*(x)$, and according to Theorems 1 and 2, $H_j^*(x) = R_j^*(x) = 0$. Then

$$\sigma \cdot e^{-\frac{\beta b}{u_j} E_j^*(x)} + E_j^*(x) = \sigma \tag{44}$$

Let $\alpha_j = (1/\sigma) \cdot E_j^*(x)$. As $E_j^*(x) \in [0,\sigma]$, $\alpha_j \in (0,1)$. Eq. (44) can be rewritten as

$$e^{-\gamma_j \alpha_j} + \alpha_j = 1 \tag{45}$$

For simplicity, denote the function $\varphi(\alpha_j)$ as

$$\varphi(\alpha_j) = e^{-\gamma_j \alpha_j} + \alpha_j - 1 \tag{46}$$

Note that

$$\varphi'(\alpha_j) = d\varphi(\alpha_j)/d\alpha_j = -\gamma_j e^{-\gamma_j \alpha_j} + 1 \tag{47}$$

According to Eq. (47), $d\varphi(\alpha_j)/d\alpha_j \big|_{\alpha_j=0} = -\gamma_j + 1 < 0$ for $\gamma_j > 1$. As $\varphi(0) = 0$, this implies that $\varphi(\alpha_j) < 0$ for $\alpha_j$ sufficiently close to 0. Note that $\varphi(1) = e^{-\gamma_j} > 0$. Hence, there must exist a solution to Eq. (45) for $\alpha_j \in (0,1)$. The second-order derivative of $\varphi(\alpha_j)$ with respect to $\alpha_j$ is $\varphi''(\alpha_j) = d^2\varphi(\alpha_j)/d\alpha_j^2 = (\gamma_j)^2 e^{-\gamma_j \alpha_j} > 0$. Thereby, $\varphi(\alpha_j)$ is a convex function. There exists at most two solutions for $\varphi(\alpha_j) = 0$. $\alpha_j = 0$ is a solution to $\varphi(\alpha_j) = 0$. This implies that there exists a unique solution to $\varphi(\alpha_j) = 0$ for $\alpha_j \in (0,1)$. Let $\alpha_j^*$ be the corresponding solution. We have $E_j^*(x) = \sigma \cdot \alpha_j^*$. Theorem 3 is proved.

Recall $I_j^*(x) = H_j^*(x) + R_j^*(x) + E_j^*(x)$, and $H_j^*(x) = R_j^*(x) = 0$. We have the following corollary.

**Corollary 1**: The asymptotic density of informed vehicles is $I_j^*(x) = E_j^*(x) = \sigma \cdot \alpha_j^*$.

According to Eq. (34), $\alpha_j^*$ is determined only by the value of $\gamma_j$ which equals $\beta b \sigma / u_j$. Thereby, when the communication frequency ($\beta$), the parameter $b$ in the communication kernel and the initial density of equipped vehicles ($\sigma$) are fixed, the service rate $u_j$ can be leveraged by transportation operators to propagate the specific information in information class $j$ to control the proportion of vehicles that can receive the specific information. Let $P_{I_j}^*$ be the information spread (i.e., proportion of informed vehicles) for the specific information packet of interest in information class $j$; we have $P_{I_j}^* = \sigma \cdot \alpha_j^* / \sigma = \alpha_j^*$.

The following theorem discusses the existence of the IFPW.

**Theorem 4** (conditions for existence of IFPW): The IFPW does not exist when $\gamma_j < 1$.

*Proof*: Note for $\alpha \in [0,1]$,
$$\varphi'(\alpha_j) = d\varphi(\alpha_j)/d\alpha_j = -\gamma_j e^{-\gamma_j \alpha_j} + 1 > -\gamma_j + 1 > 0 \qquad (48)$$
Thereby, $\varphi(\alpha_j)$ increases monotonically with respect to $\alpha_j$ for $\alpha_j \in (0,1)$. As $\varphi(0) = 0$, there exists no solution to $\varphi(\alpha_j) = 0$ for $\alpha_j \in (0,1)$. This implies that the vehicles are far from the location (location is labeled as $x$) where the information packet is generated, and the asymptotic solution of $E_j(x,t)$ is 0. As all informed vehicles will become information-excluded vehicles, this result indicates that no vehicle can receive the specific information packet of interest of information class $j$ if they are far from the location where the information is generated in this case. Thereby, the IFPW does not exist for $\gamma_j < 1$.

Theorem 4 shows that if the initial density of equipped vehicles and the service rate are high, the specific information packet of interest can only be propagated locally. Vehicles that are far from the location where the specific information is generated cannot receive it. This property can be used by transportation operators to design effective control strategies to propagate information within a small vicinity (e.g., sudden braking information, lane merge information). It should be noted that $u_j$ will impact the propagation distance of the specific information of interest when the IFPW does not exist. Through a numerical example, we will show that if $u_j$ is set such that $\gamma_j$ is closer to 1 ($\gamma_j < 1$), the specific information of interest will be propagated further away. If $\gamma_j > 1$, the specific information packet will form a wave to be propagated in the network.

As discussed earlier, we cannot derive an analytical solution for the asymptotic IFPW speed even if it exists. The IFPW speed will be computed using the numerical method introduced in the next section. Note that the asymptotic IFPW speed is significantly impacted by the queuing delay which is determined by the two control parameters (i.e., $n_j$ and $u_j$) simultaneously. To meet the application needs of information in an arbitrary information class $j$ related to information spread, time delay to reach a target location and spatial coverage, the values for the two control parameters $n_j$ and $u_j$ can be determined as follows: first, choose $u_j$ appropriately according to Theorem 4 if the information needs to be only propagated locally. If the information needs to be propagated in the network, then determine $u_j$ appropriately according to Theorem 4 and corollary 1 so that information spread can be satisfied. Third, determine $n_j$ appropriately to control the IFPW speed so that it can reach the target location in the desired time.

**4.2 Numerical solution method**

The analytical solutions for the various information classes introduced in previous section only apply to homogeneous traffic conditions. Under heterogeneous conditions, the IFPW may not be stable due to the non-uniform impact of traffic flow dynamics on information dissemination. To analyze how information is spread in space and time under heterogeneous conditions, this section proposes a numerical solution method based on Kim et al. (2017) to solve the two-layer model. The numerical solution method helps to: (1) estimate the IFPW speed under both homogeneous and heterogeneous traffic conditions, and (2) estimate the distance the specific information can be propagated when the IFPW does not exist, under both homogeneous and heterogeneous traffic conditions, and (3) estimate the density of informed vehicles under heterogeneous conditions.

The numerical solution method discretizes space and time into cells of length $\Delta x$ and time interval $\Delta t$, respectively. Let $1, 2, 3 \cdots$ denote the cells in the highway sequentially. The fourth-order Runge-Kutta method will be used to approximate the densities of vehicles by vehicle class (i.e., $S_j(x, t), H_j(x, t), R_j(x, t)$ and $X_j(x, t)$) changed according to the SHRE model in the upper layer. To solve the LWR model in the lower layer, the generalized cell transmission finite difference method proposed by Daganzo (1995) is used to approximate Eq. (24) and Eq. (25) as follows

$$[k(x, t + \Delta t) - k(x, t)]/\Delta t = [q(x - \Delta x, t) - q(x, t)]/\Delta x \tag{49}$$

$$q(x, t) = \min\left\{T(k(x, t)), Q\left(k_{jam} - k(x + \Delta x, t)\right)\right\}, \tag{50}$$

where $T$ specifies the maximum flow that can be sent by the upstream cell and $Q$ specifies the maximum flow that can be received by the downstream cell. $k_{jam}$ is the jam traffic density. Let $U$ denote the unequipped vehicles. The steps to solve the two-layer model numerically are as follows:

Step 1: At time 0 ($t = 0$), obtain the initial number of vehicles of each class $z, z \in \{S_j, H_j, R_j, E_j, U\}$ and corresponding density of vehicles of each vehicle class in each cell. Let $t = t + \Delta t$.

Step 2: Solve the lower-layer model to determine the flow in each cell $x$ (i.e., $q(x, t)$) that advances to the downstream cell according to Eq. (49) and Eq. (50). Update the number of vehicles in each cell.

Step 3: Calculate the number of vehicles of each class $z \in \{S_j, H_j, R_j, E_j, U\}$ that advance to the downstream as follows:

$$q_z(x, t) = \frac{k_z(x, t - \Delta t)}{k(x, t - \Delta t)} \cdot q(x, t), \quad z \in \{S_j, H_j, R_j, E_j, U\}, \tag{51}$$

where $q_z(x, t)$ is the traffic flow of class $z$ leaving cell $x$ at time interval $t$ and $k_z(x, t - \Delta t)$ is the density of class $z$ in cell $x$ and time $t - \Delta t$.

Step 4: Update the density of vehicles by vehicle class in each cell of the upper layer using the discrete multiclass flow conservation law, as follows:

$$[k_z(x, t) - k_z(x, t - \Delta t)]/\Delta t = [q_z(x - \Delta x, t) - q_z(x, t)]/\Delta x, z \in \{S_j, H_j, R_j, E_j, U\}. \tag{52}$$

where $k_{S_j}(x, t), k_{H_j}(x, t), k_{R_j}(x, t)$ and $k_{E_j}(x, t)$ represents $S_j(x, t)$, $H_j(x, t)$, $R_j(x, t)$ and $E_j(x, t)$, respectively that describe the density of vehicles of each class in the upper layer.

Step 5: Approximate the density of vehicles by vehicle class (i.e., $S_j(x, t), H_j(x, t), R_j(x, t)$ and $E_j(x, t)$) that are changed according to the SHRE model in the upper layer.

Step 6: If the predetermined time length is reached, then stop. Otherwise, let $t = t + \Delta t$, and go to Step 2.

The numerical method solves the discretized LWR model and the SHRE model sequentially to capture the effects of traffic flow dynamics on information dissemination. It is worth noting that to reduce computational load, the convolution term ($\int_\Omega R_j(y, t) \cdot K(x, y) dy$) in Eq. (23) can be approximated using the Fast Fourier Transform (FFT) method. More details of the numerical method and the FFT method can be found in Kim et al. (2017). The numerical method can provide the density of vehicles by vehicle class at each cell and time interval.

The numerical method can also be used to verify the analytical solutions of density of informed vehicles and to approximate the IFPW speed under both homogeneous and heterogeneous traffic conditions. Note that IFPW consists of two waves: the forward wave which travels in the direction of vehicular traversal and the backward wave which travels opposite to the direction of vehicular traversal. Correspondingly, there exist two IFPW speeds, the forward and backward IFPW speeds. The method to estimate the two IFPW speeds is as follows. Let $t_1$ and $t_2$ be two arbitrary time intervals. Without loss of generality, let $t_1 > t_2$. Let $z_{j,0}$ be the density of vehicles in an arbitrary vehicle class $z_j$. $z_{j,0}$ is set between the minimum and maximum density of vehicle class $z_j$. Then, at the two time intervals $t_h, h = 1,2$, there exist two cells on the two wave fronts, respectively, for which the density of vehicle class $z_j$ is most close to $z_{j,0}$. Denote the two cells as $l_{B,t_h}$ and $l_{F,t_h}$ ($h = 1,2$), respectively. Without loss of generality, let $l_{B,t_h}$ be the cell in the backward IFPW and the $l_{F,t_h}$ be the cell in the forward IFPW. Then, the forward IFPW (labeled as $c_F$) and the backward IFPW (labeled as $c_B$) can be approximated as

$$c_F = \frac{(l_{F,t_2} - l_{F,t_1})\Delta x}{t_1 - t_2} \tag{53}$$

$$c_B = \frac{|(l_{B,t_2} - l_{B,t_1})\Delta x|}{t_1 - t_2} \tag{54}$$

## 5. Numerical experiments

This section discusses several numerical experiments to illustrate the application of the proposed method to control multiclass information flow propagation. Consider a highway with 30 km length. Discretize the highway uniformly into 2000 cells. Table 1 shows other inputs in the experiment.

Table 1. Experiment parameters

| Traffic flow parameters | Value |
| --- | --- |
| Free flow speed ($u_f$) | 108 km/h |
| Time interval ($\Delta t$) | 0.5 seconds |
| Cell length ($\Delta x$) | 15 meters |
| Number of lanes | 1 |
| Market penetration rate ($W$) | 50% |

### 5.1 Calibration of the communication kernel

To calibrate the communication kernel function in Eq. (26), NS-3 will be used to simulate the success rate of one-hop V2V communications under different traffic flow densities. NS-3 is a discrete network simulator that can simulate and test a spectrum of communication protocols efficiently. Recently, NS-3 has been used to simulate V2V communications and evaluate the performance of communication protocols for vehicular ad hoc networks (see e.g., Dey et al., 2016; Noori and Olyaei., 2013; Talebpour et al., 2016). The inputs for the V2V communication related parameters in NS-3 are shown in Table 2. The simulation is operated based on IEEE 802.11p protocol in 5.9 GHz band with channel capacity 3 Mbps and communication power 500 m. In V2V communications, whether a receiver vehicle can successfully receive an information packet from a sender vehicle is primarily decided by two factors: the reception signal power, and the noise and interference. The reception signal power determines whether the receiver vehicle can receive the signals from the sender vehicles, and the level of noise and interference determines the probability of reception error. In this simulation, the Friis propagation loss model (Benin et al., 2012) is used in NS-3 to calculate the reception signal power. It characterizes the impacts of transmission power, distance between receiver and sender, transmission gain, and reception gain on reception signal power. The receiver vehicle receives the information packet only if the reception signal power is larger than the energy detection threshold $-96$ dBm. To estimate the noise and interference, the signal to (interference and) noise ratio (SINR) model is used in NS-3 simulation. The SINR is the ratio of the power of a certain signal of interest over the sum of the interference power (from all the other interfering signals) and the power of some background noise (for details, see Wang et al. 2018). The threshold of SINR is set as 5 dB (Hisham et al., 2016; Hisham et al., 2017), indicating that the V2V communication is considered to be successful if SINR is larger than 5 dB; otherwise, it will be considered as a communication failure.

Table 2. Inputs for NS-3 parameters

| Parameters | Value |
| --- | --- |
| IEEE 802.11p channel capacity | 3 Mbps |
| Band | 5.9 GHz |
| Communication frequency | 2 Hz |
| Communication power/distance | 500 m |
| Minimum contention window | 15 slots |
| Energy detection threshold | -96 dBm |
| Noise floor | -99 dBm |
| SINR threshold | 5 dB |

Table 3. Maximum number of communication servers, and calibrated parameters in communication kernel using NS-3 simulation

| Density (veh./km) | 10 | 20 | 30 | 40 | 50 | 60 | 70 | 80 | 90 | 100 |
| --- | --- | --- | --- | --- | --- | --- | --- | --- | --- | --- |
| $a$ | 0.362 | 0.351 | 0.313 | 0.292 | 0.267 | 0.258 | 0.216 | 0.199 | 0.176 | 0.153 |
| $b$ | 0.621 | 0.576 | 0.531 | 0.499 | 0.434 | 0.392 | 0.357 | 0.291 | 0.268 | 0.243 |
| $N_{max}$ | 125 | 62 | 41 | 31 | 25 | 21 | 18 | 15 | 14 | 12 |

Recall that all equipped vehicles within communication range are assumed to share the bandwidth equally. Suppose the single unit of an information packet is 500 bytes. To prevent information congestion effects that would occur if the channel capacity is full, the maximum number of communication servers can be calculated as follows

$$N_{max} = \left[\frac{C}{2 \cdot k \cdot \beta \cdot R \cdot W \cdot \chi}\right].$$

where $R$ is the communication range, $C$ is the channel capacity (3 Mbps), $W$ is the market penetration rate of V2V-equipped vehicles (50% in this study). $\chi$ is the information packet size (500 bytes), and $k$ is the traffic flow density. The operator $[h]$ means the largest integer less than $h$. The calculated $N_{max}$ values for different traffic flow densities are shown in Table 3. They will be used as the total number of communication servers under the corresponding traffic flow densities. To account for the impacts of positions of vehicles on success rate of V2V communications, vehicles are assumed to be randomly distributed along the 30 km highway. The simulation is conducted for 30 minutes, and is repeated 100 times. The calibrated parameters in the communication kernel are presented in Table 2. Figure 3 illustrates the simulated success rate of one-hop propagation and calibrated communication kernels at $k = 40$ veh/h and $k = 60$ veh/h. The R-squared values of the calibrated communication kernel at $k = 40$ veh/h and $k = 60$ veh/h are 0.96 and 0.94, respectively, indicating that the calibrated communication kernel robustly captures the relationship between success rate of one-hop propagation and the distance of the sender vehicle to the receiver vehicle. In addition, for the same distance, the success rate of one-hop propagation at $k = 60$ veh/h is less than it is at $k = 40$ veh/h. This because the communication interference increases if more vehicles are located within the communication range of a sender vehicle, causing greater communication failure.

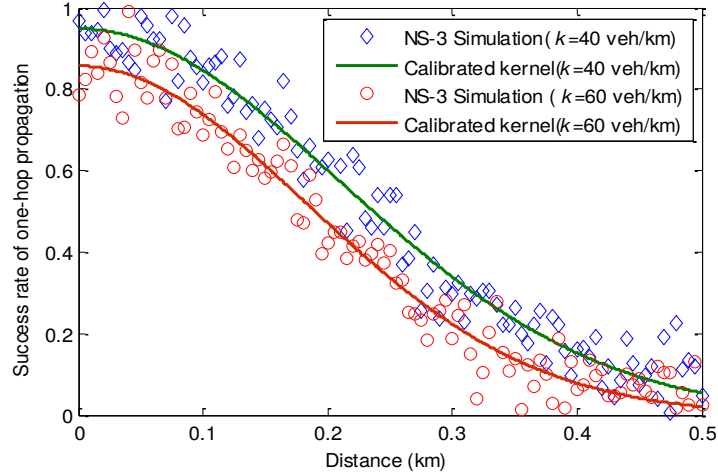

**Figure 3**. Calibrated communication kernel for $k = 40$ veh/h and $k = 60$ veh/h

### 5.2 IFPW under homogeneous conditions

*5.2.1 Asymptotic density of informed vehicles and IFPW speed*

The following example shows how to calculate the asymptotic density of informed vehicles analytically under homogeneous conditions. Suppose the traffic flow density is 40 veh/h, and the market penetration rate of equipped vehicles is 50%. Then, for each cell, the number of equipped vehicles is 0.3 veh/cell ($\sigma$). According to Table 3, the parameters $a$ and $b$ in Eq. (26) are 0.292 and 0.499, respectively. Suppose the specific information of interest is from information class $j$. Assume the number of communication servers ($n_j$) assigned to send the information packets in information class $j$ is 11 and the corresponding mean communication service rate $u_j$ is 0.05 packet/seconds (i.e., mean service time is 20 seconds). Note $n_j u_j - \lambda_j = 11 \times 0.05 - 0.5 = 0.05 > 0$ and $\gamma_j = \beta b \sigma / u_j = 2 \times 0.499 \times 0.5 * \frac{0.6}{0.05} = 2.994 > 1$. According to Theorem 4 and Corollary 1, the asymptotic density of informed vehicles exists. Corollary 1 indicates that $I_j^*(x) = 0.3 \cdot \alpha_j^*$, where $\alpha_j^*$ is the unique solution of the nonlinear solution $e^{-3.384\alpha_j} + \alpha_j - 1 = 0$ for $\alpha_j \in [0,1]$. Using Newton method to solve the nonlinear equation, we have $\alpha_j^* = 0.9613$. Thereby, the asymptotic density of informed vehicles is $I_j^*(x) = 0.3 \cdot 0.9613 = 0.288$ vehicel/cell $= 19.2 \ veh/km$. The information spread (proportion of informed vehicles) is $P_{I_j}^* = \alpha_j^* = 0.9613$.

Figure 4 compares the information spread of the specific information packet of information class $j$ under different traffic flow densities at $n_j = 11$ and $u_j = 0.05$ packets/second. It shows that the numerical solutions overlap with the analytical solutions, implying that the numerical algorithm proposed in Section 4.2 can effectively solve the two-layer model. Figure 4 also demonstrates that when traffic flow density increases, the information spread also increases as more vehicles will propagate it in an unit of time. This indicates that under higher traffic flow density scenarios, the mean communication service rate of information packets of class $j$ can be reduced for the same information spread.

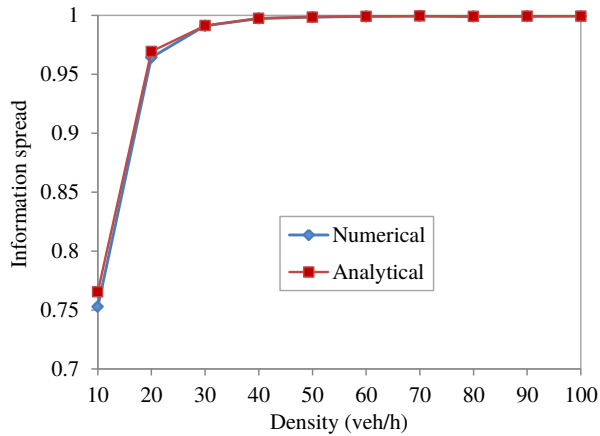

**Figure 4** Information spread at $n_j = 11$ and $u_j = 0.05$ under different traffic densities

Suppose the traffic flow density is 40 veh/h and the specific information packet of interest in information class $j$ is generated by a vehicle at time 0 and location 0. Let $\lambda_j = 0.5$ packets/second, $n_j = 20$ and $u_j = 0.03$ packets/second. Figure 5 shows the spatial distribution of density of vehicles by vehicle classes at $t = 150$ seconds and $t = 230$ seconds. It indicates that the IFPW can form the same shape to move forward and backward. Most of the information-holding and information-relaying vehicles are located close to the wave front. This is because when $\lambda_j < n_j u_j$ and $u_j > 0$, the information packet in information class $j$ will experience finite queuing delay and communication service time. Thereby, the vehicles that receive the specific information packet of interest a long time ago will exclude it from the system. The information-holding and information-relaying vehicles will become information-excluded vehicles eventually.

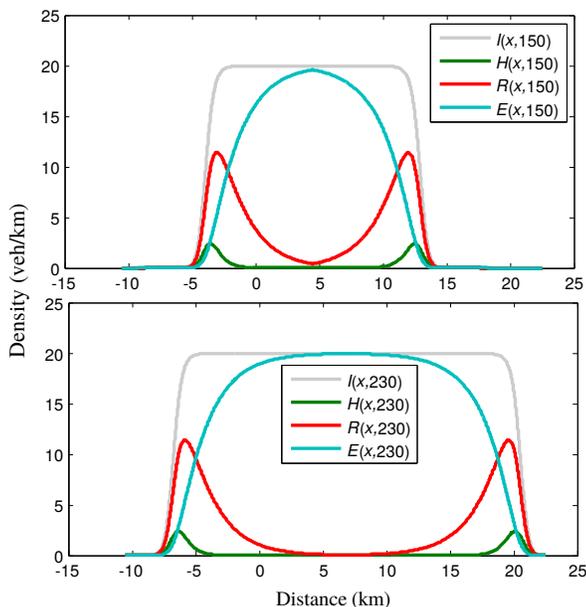

**Figure 5**. Density of vehicles by vehicle class at $t = 150$ seconds and $t = 230$ seconds

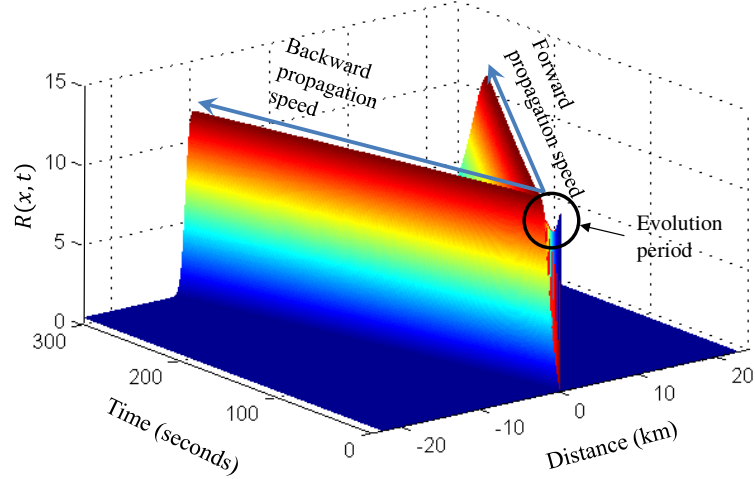

**Figure 6**. Density of information-relaying vehicles in space and time at $k = 40\ veh/km$

To analyze the asymptotic IFPW speed, Figure 6 shows the spatiotemporal distribution of density of information-relaying vehicles. It illustrates that the specific information packet of interest of information class $j$ is propagated backward and forward at a uniform speed reached only a few seconds after it is generated. To numerically estimate the IFPW speed, Figure 7 shows the spatial distribution of density of information-excluded vehicles at $t = 150$ seconds and $t = 230$ seconds. Let $R_0 = 10\ veh/km$ be the reference density in Eq. (53) and Eq. (54). According to Figure 7, the cells whose densities of information-relaying vehicles in the backward and forward IFPWs are most close to $R_0$ at $t = 150$ seconds are located at $-2.085$ km and $11.055$ km, respectively. The cells whose densities of information-relaying vehicles in the backward and forward IFPWs are most close to $R_0$ at $t = 230$ seconds are located at $-4.875$ km and $18.645$ km, respectively. According to Eq. (53) and Eq. (54), the forward and backward IFPW speeds can be estimated as:

$$c_F = \frac{(l_{F,t_2} - l_{F,t_1})\Delta x}{t_1 - t_2} = \frac{-2.085 + 4.875}{150 - 230} = 0.09488 km/s = 341 km/h$$

$$c_B = \frac{|(l_{B,t_2} - l_{B,t_1})\Delta x|}{t_1 - t_2} = \frac{|-2.085 + 4.875|}{150 - 230} = 0.03487 km/s = 125 km/h$$

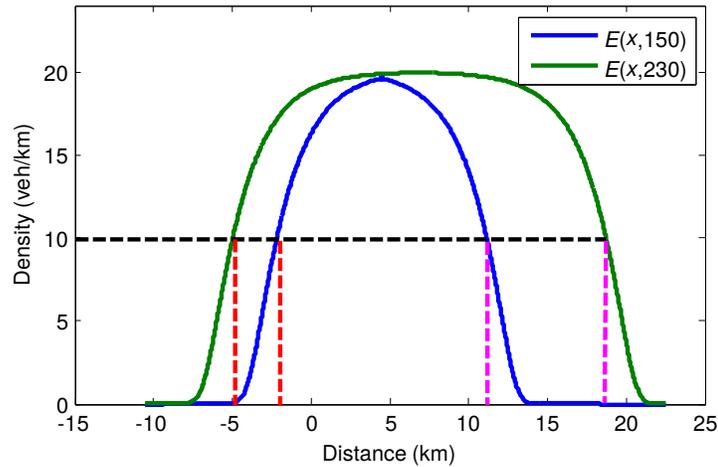

**Figure 7**. Density of information-excluded vehicles at $t = 150$ seconds and $t = 230$ seconds

*5.2.2 Scenarios where the information packet can be propagated only locally*

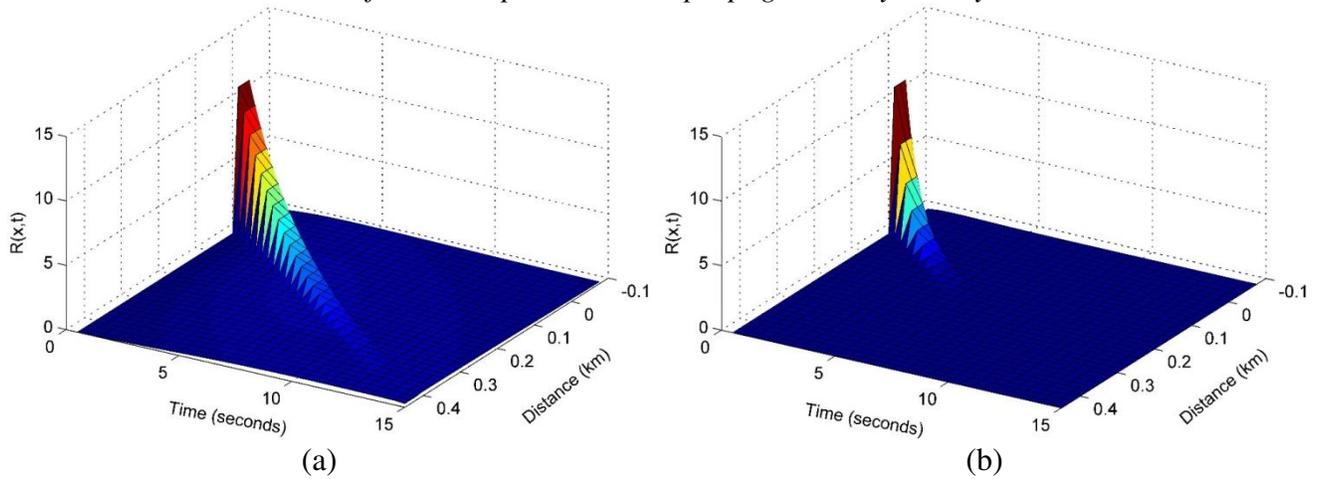

**Figure 8**. Scenarios for which the information packets are propagated only locally: (a) Density of information-relaying vehicle in space and time for $n_j = 20$ and $u_j = 0.323$; (b) Density of information-relaying vehicle in space and time for $n_j = 20$ and $u_j = 0.9$.

As Theorem 4 indicates, the IFPW does not exist when the mean communication service rate is high enough such that $\gamma_j < 1$. In this case, the specific information packet in information class $j$ can be propagated only locally. This property can be leveraged to send information packets in a small vicinity of where they are generated. The following example seeks to demonstrate how to control the propagation distance by leveraging communication service rate when information is propagated locally. Suppose the traffic flow density is 40 veh/h, the average arrival rate of information packets of information class $j$ is 2 packets/second, and the number of assigned communication servers for information class $j$ is 20. Let the density of information-relaying vehicles at location 0 and time 0 be 11 veh/km and 0 elsewhere. Figure 8(a) and Figure 8(b) show the spatiotemporal distribution of density of information-relaying vehicles at $u = 0.323$ and $u = 0.9$, respectively. Note $\gamma_j < 1$ in both cases. Figure 8 illustrates that the density of information-relaying vehicles decreases to 0 in space and time. Recall only information-relaying vehicles can propagate the specific information packets of interest. This implies that the specific information packet can only be propagated locally. Vehicles far away from location 0 where the specific information packet is generated will not receive it. It can be noted that the density of information-relaying vehicles decreases to 0 at 350 meters and 150 meters downstream of location 0 at $u = 0.323$ and $u = 0.9$, respectively. This implies the information packets can be propagated further away under a lower communication service rate. Thereby, the mean communication service rate can also be leveraged to propagate information to different distances.

*5.2.3 Integrated impacts of $n_j$ and $u_j$ on asymptotic IFPW speed and density of informed vehicles*

Suppose the traffic flow density is 50 veh/km, and the arrival rate of the information packets in information class $j$ is 1 packet/second. To analyze the impacts of $n_j$ and $u_j$ on IFPW speed and density of informed vehicles, $u_j$ is varied from 0.05 packets/second to 0.25 packets/second. According to M/M/$n_j$ queuing theory, to enable propagation of the specific information packet of information class $j$, $\lambda_j/(n_j u_j)$ must be less than 1. Thereby, the minimum number of communication servers assigned to send information packets of information class $j$ are 21, 11, 7, 6, 5 for $u_j = 0.05, 0.1, 0.15, 0.2, 0.25$, respectively. Figure 9 shows the asymptotic forward IFPW speed for various values of $n_j$ and $u_j$. It shows that when $u_j$ is fixed, the asymptotic forward IFPW speed increases monotonically with respect to $n_j$. This is because as more communication servers are assigned to information class $j$, the mean waiting time of the specific information packet in the queue will reduce (see Figure 10). Thereby, it can be transmitted faster by the informed vehicles. Figure 9 also shows that for a fixed $n_j$, the IFPW speed decreases monotonically with respect to mean communication service rate $u_j$ in most cases because an increase in mean communication service rate will reduce the transmission duration of the packet. However, in some cases (e.g., $n_j = 22, 23$, etc.), increase in $u_j$ may decrease the forward IFPW speed. This is because for a fixed $n_j$, increase in $u_j$ will increase the mean waiting time in the queue (see Figure 10). Thereby, unlike that of the number of communication servers, the effect of mean communication service rate on the IFPW speed is more intricate. The proposed method in this study aids in determining the appropriate mean communication service rate for each information class to satisfy its application needs in terms of propagation performance.

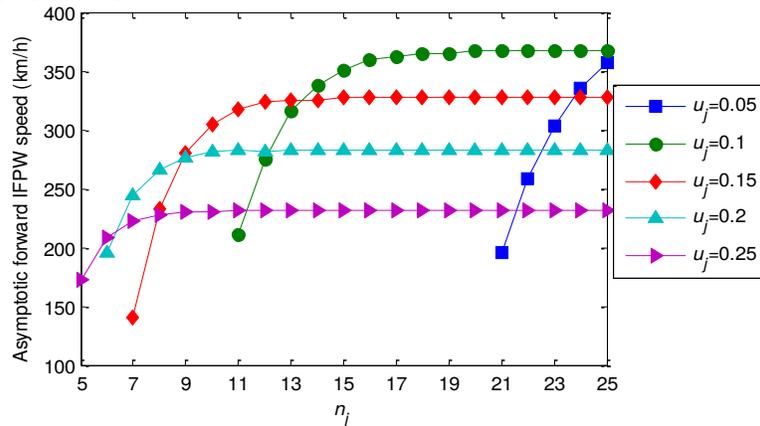

**Figure 9**. Impacts of $n_j$ and $u_j$ on asymptotic forward IFPW speed of an information packet of information class $j$

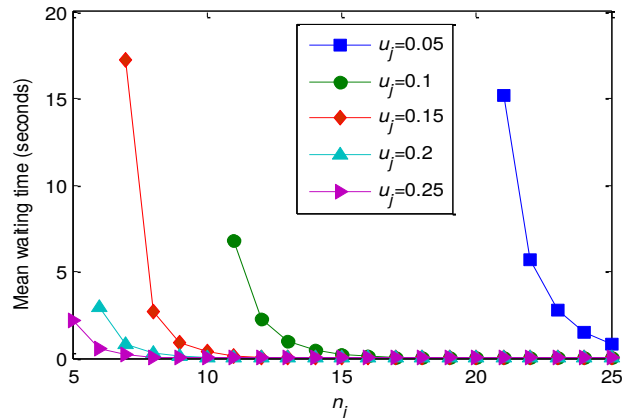

**Figure 10**. Mean waiting time of information packets in the queue for various values of $n_j$ and $u_j$

Figure 11 shows the information spread of the specific information of interest under different communication service rates. As the value of $u_j$ increases, information spread decreases monotonically, implying that less number of vehicles will receive the specific information of interest. This is because an informed vehicle will exclude the specific information packet of interest from the communication servers faster under higher mean communication service rate. It is worth mentioning that the number of communication servers has no effect on information spread. Thereby, to design effective control strategies for propagating information packets in different classes, the mean communication service rate can be determined first to obtain the desired information spread. Then, the appropriate number of communication servers can be determined to be assigned to different information classes to control their propagation speed. Further, Figure 11 shows that the numerical solutions are almost identical to the analytical solutions.

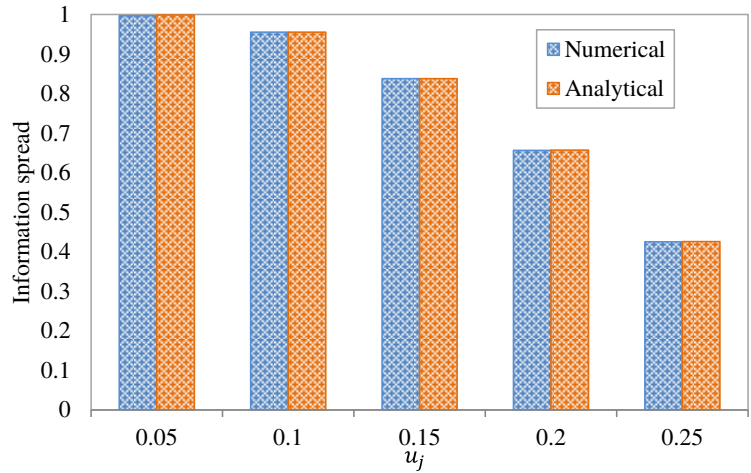

**Figure 11**. Comparison of numerical and analytical solutions of information spread for different values of $u_j$

**5.3 Control of multiclass information flow propagation under homogeneous and heterogeneous traffic conditions**

*5.3.1 Control of multiclass information flow propagation under homogeneous traffic conditions*

This section analyzes the control of multiclass information propagation to meet application needs of various classes simultaneously. Recall that the queuing system for each information class is independent. Thereby, the number of communication servers and the mean communication service rate can be controlled for each information class to achieve desired propagation performance related to information spread (related to density of informed vehicles), time delay bounds (related to IFPW speed) and spatial coverage (related to existence of IFPW). Suppose traffic flow density is 50 veh/km and information from three information classes (labeled information class 1, 2 and 3) is propagated over the traffic stream. Let information class 1 contain "urgent" information (e.g., traffic accident blocks the freeway link fully). It is desirable for this information to reach all upstream and downstream vehicles with low latency. Information class 2 is constituted by less urgent information; for example, routing information. It is delay-tolerant and is expected to reach a lower proportion of equipped vehicles compared to information class 1 to avoid congestion in other routes. Information class 3 contains information with limited impact area, which needs to be propagated locally, for example, information of sudden braking of a vehicle, lane merge information, etc. Suppose the mean arrival rate of information packets of information class 1, 2 and 3 are 0.3, 0.8 and 1.2 packets/second, respectively. According to Table 3, the total number of communication servers that can be assigned under traffic flow density 50 veh/km is 25. Let the number of communication servers assigned to information classes 1, 2 and 3 be 12, 8 and 5, respectively. The mean communication service rates are set correspondingly as $u_1 = 0.05$ packets/second, $u_2 = 0.2$ packets/second, and $u_3 = 0.4$ packets/second. Note that $\gamma_1 > 1, \gamma_2 > 1$, and $\gamma_3 < 1$. According to Theorem 4, the IFPW exists for propagation of information packets of information classes 1 and 2 while it does not exist for information packets of class 3.

Figure 12 compares the forward and backward IFPW speed of information classes 1 and 2. It shows that both forward and backward IFPW speeds of information class 1 are greater than those of information class 2. In addition, the proportion of vehicles (information spread) informed with the packets of information class 1 and information class 2 are 99.8% and 65.6%, respectively. Thereby, under the designed control strategy, packets from information class 1 can reach more number of vehicles with lower time delay compared to packets from information class 2. Figure 13 shows the contour of the density of information-relaying vehicles. It indicates that vehicles relaying the specific information packet of information class 3 decreases dramatically with space and time. The specific information packet is almost excluded by all vehicles beyond the locations 480 meters downstream and 200 meters upstream of its point of origin (i.e., location 0). Thereby, the information packets of class 3 are only propagated to a small area.

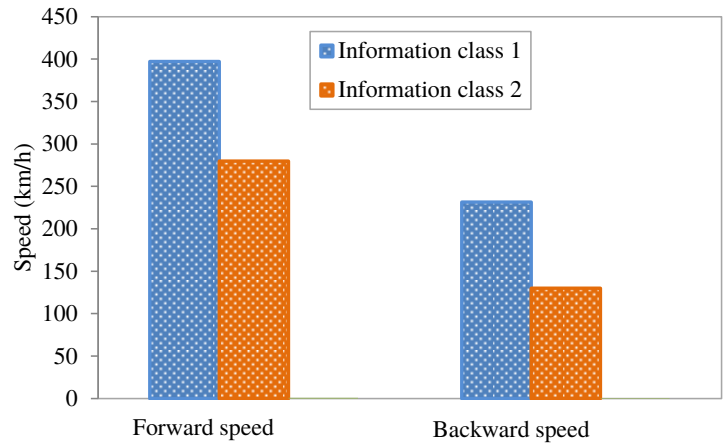

Figure 12. Comparison of forward and backward propagation speeds of information classes 1 and 2

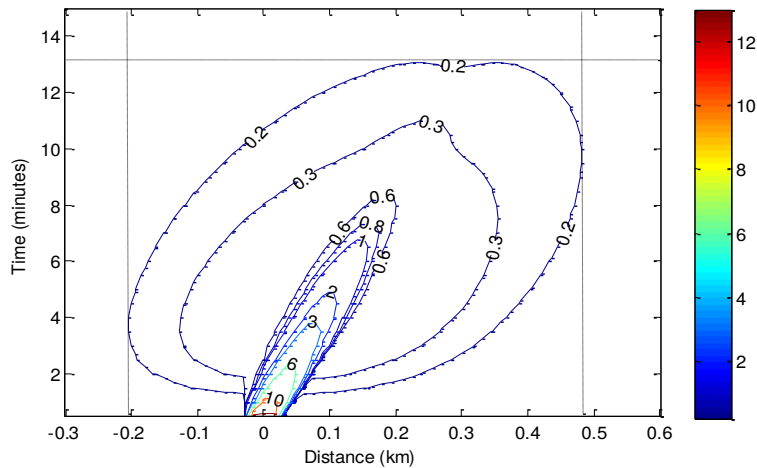

Figure 13. Contour of density of information-relaying vehicles of information class 3

*5.3.2 Control of multiclass information flow propagation under heterogeneous conditions*

This section address the control of information flow propagation under heterogeneous conditions. Similar to Kim et al. (2017), consider that a traffic accident happens at time 0 on a unidirectional highway with a traffic flow density of 50 veh./km. As illustrated by Figure 14, the incident occurs at location A. It reduces the link capacity by one third for 4 minutes before it is cleared. The congested traffic and the free flow traffic departing from the incident occurrence location are separated by Line AB. The occurrence and clearance of the incident induce two forward propagating traffic waves denoted by lines AD and BF, respectively. After the incident occurs, vehicles are jammed at the incident location, leading to a traffic wave propagating backward.

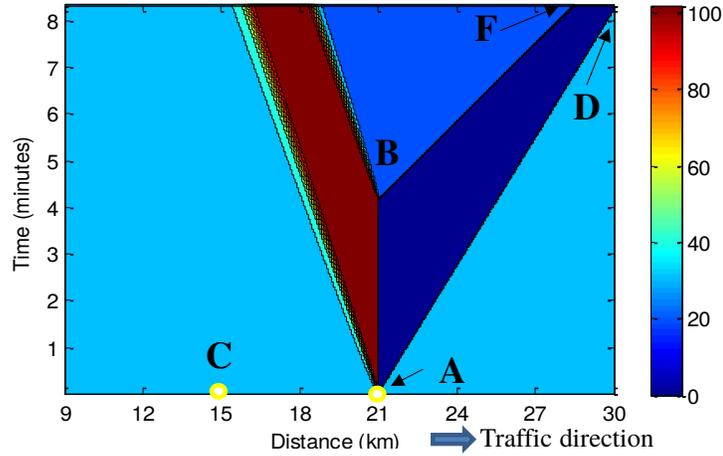

Figure 14. Contour of traffic density

Suppose information packets of three different information classes are generated at location C, and their arrival rates are identical. We label them information classes 1, 2, and 3. Information classes 1 and 2 contain routing information and are expected to reach the same number of equipped vehicles. However, information packet of class 2 is expected to be propagated faster than information packet of class 1 as it contains information related to the traffic accident, which requires more imminent response from the vehicles. Information class 3 contains information related to the level of traffic congestion induced by the traffic accident. Hence, information packets of class 3 are expected to be received by all vehicles in the impacted area.

To achieve these objectives, let the number of communication servers assigned for information classes 1, 2 and 3 be 5, 10 and 10, respectively. The mean communication service rates for the three information classes are $u_1 = 0.15$ packets/second, $u_2 = 0.15$ packets/second, and $u_3 = 0.06$ packets/second. The numerical solution method will be used to calculate the information propagation speed and the proportion of informed vehicles (information spread) for the three information classes.

Figures 15(a) and 15(b) compare the backward and forward IFPW speeds of information classes 1 and 2, respectively. They illustrate that information packets in both classes are propagated very fast in the uncongested area which is not impacted by the traffic accident (see stages $a$ and $a'$ in Figures 15(a) and (15b), respectively). The propagation speed decreases significantly when the information packets arrive at the congested area induced by the traffic accident (stages $b$ and $b'$ in Figures 15(a) and (15b), respectively). This is because higher traffic density of vehicles can increase communication interference, causing significant communication failures. The IFPW speed is recovered to the original value when the congested area is passed to catch up with the normal traffic (stages $c$ and $c'$ in Figures 15(a) and (15b), respectively). It is important to note that information packets in information class 2 are propagated faster than those in information class 1. For example, the information packets of information class 1 take about 4 minutes and 5 minutes to reach the points G and H located at 9 km and 30 km, respectively. In comparison, it only takes 2 minutes 40 seconds and 4 minutes for information packets of information class 2 to reach the two locations, respectively. These results indicate that under heterogeneous conditions, controlling the number of communication servers assigned to each information class can significantly impact the time delay of the information packets to reach the targeted locations. Figures 15(a) and 15(b) also reveal that the information spread (i.e., $P_{I_j}^*$) is the same in space and time. This implies that the number of communication servers only impacts the propagation speed, but not the asymptotic proportion of informed vehicles.

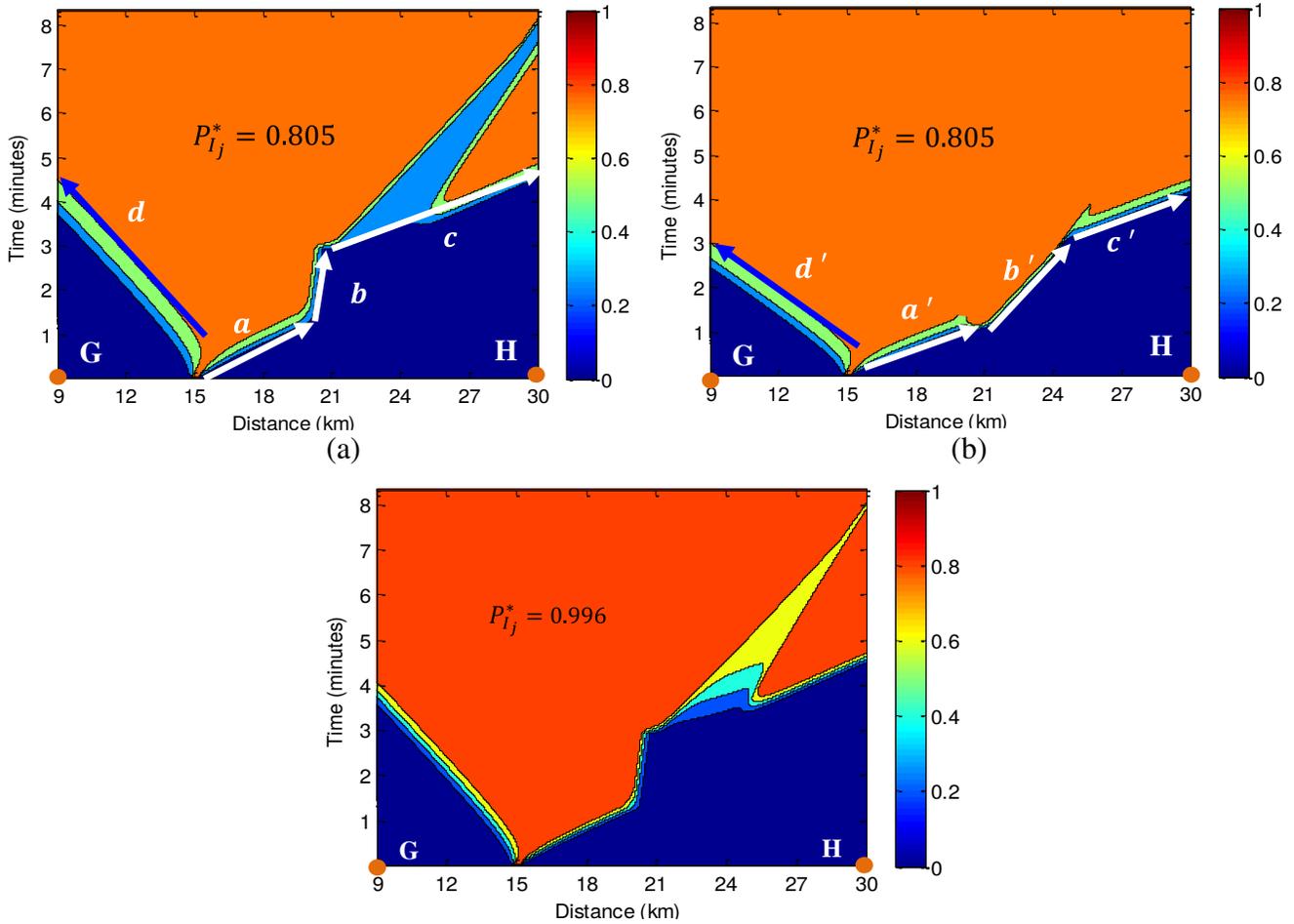

Figure 15. Contours of proportion of information-excluded vehicles of information packets of information classes 1, 2 and 3: (a) information class 1; (b) information class 2; (c) information class 3

Figure 15(c) shows that $P_{I_3^*} = 0.996$, implying that almost all equipped vehicles can receive information packets of class 3. Note that the number of assigned communication servers for information classes 2 and 3 are identical. Figures 15(b) and Figure 15(c) indicate that smaller mean communication service rate will enable more number of vehicles to receive the specific information packet in corresponding information class. However, it may reduce the forward and backward propagation speeds. Figure 15(c) shows that it takes longer time for information packets in information class 3 to be delivered at locations G and H, compared to those of information class 2. This is because reducing the mean communication service rate will increase the mean waiting time of information packets in the queue. Thereby, under the designed scenarios, the information propagation speed is reduced due to increased mean waiting time. The proposed method in this study can aid transportation operators to determine the mean communication service rate and the number of communication servers assigned to each information class to control the information propagation performance under both homogeneous and heterogeneous traffic conditions.

# 6. Conclusions

The traffic information propagated in a V2V-based traffic system can be grouped into different classes based on application needs related to information spread, time delay bounds, and spatial coverage. To meet these needs of multiclass information under different traffic flow and communication environments, this study proposes a queuing strategy for equipped vehicles to propagate the received information packets. The queuing strategy enables control for multiclass information propagation by leveraging two control parameters, the number of communication servers and the mean communication service rate. The spatiotemporal propagation of information in different information classes under the designed queuing strategy is characterized by a two-layer analytical model. The upper layer is an IDE system derived to model information dissemination under the designed queuing strategy, and a LWR model is used in the lower layer to describe the traffic flow dynamics. An analytical solution of asymptotic density of informed vehicles is developed under homogeneous traffic conditions. It helps to analyze the relationship between the density of informed vehicles and the two control parameters in the queuing strategy. In addition, the necessary conditions for existence of IFPW are derived. It describes the conditions under which the specific information packets will be propagated only locally. A numerical solution is proposed to solve the two-layer model to estimate the IFPW speed, which helps to estimate the time delay for an information packet to reach the target location.

  Numerical experiments using the proposed model suggest that the mean communication service rate significantly impacts the asymptotic density of informed vehicles. Also, all else being equal, an increase in the number of communication servers assigned to an information class will increase the IFPW speed of the information packets in this information class. In addition, information will be propagated only locally under a high communication service rate because each information packet has little transmission duration. These findings provide valuable insights for controlling the propagation of multiclass information to achieve desired operational performance in a V2V-based traffic system. That is, they provide valuable tools to a traffic control center to target different information-based solutions for different traffic-related problems that arise regularly in urban areas. For example, they can be used to disseminate area-wide controlled information dissemination strategies to manage traffic conditions under a severe accident, and to simultaneously manage the impacts of a work zone by ensuring information (of a different class) is disseminated to vehicles in only a certain vicinity of it, thereby enabling vehicles to seamlessly receive and propagate multiclass information. Hence, this study can be leveraged to develop a new generation of information dissemination strategies focused on enabling specific V2V-based applications for traffic situations that arise on a daily basis.

  This study can be extended in a few directions. First, analytical solutions of the IFPW speed can be derived to provide insights on the relationship between the two control parameters in the queuing strategy and the resulting IFPW speed of information packets in each information class. Second, this study only considers control of information flow propagation in a corridor. The performance of the proposed method on control of network-level information flow propagation needs to be investigated. Third, this study assigns the received information packets into different queues according to the information classes they belong to. It assumes that the number of available communication servers is larger than the number of information classes. This assumption may not hold in scenarios of high traffic density, where the maximum number of information packets (i.e., $N_{max}$) an equipped vehicle can be transmit in one-hop propagation is small due to information congestion (Wang et al., 2018). To address this, other queuing strategies such as preemptive priority and non-preemptive priority queuing systems will be developed to control the propagation of information of different information classes.

**Acknowledgements**

This study is based on research supported by the Center for Connected and Automated Transportation (CCAT) Region V University Transportation Center funded by the U.S. Department of Transportation, Award #69A3551747105. The third author is partly supported by the Natural Science Foundation of China (71701108), and the Natural Science Foundation of Zhejiang Province (LQ17E080007). Any errors or omissions remain the sole responsibility of the authors.

**References**

Benin, J., Nowatkowski, M., Owen, H., 2012. Vehicular network simulation propagation loss model parameter standardization in ns-3 and beyond. 2012 Proceedings of IEEE Southeastcon, March, pp. 1-5, Florida, USA.

Daganzo, C. F., 1995. A finite difference approximation of the kinematic wave model of traffic flow. Transportation Research Part B: Methodological, 29 (4), 261–276.

Dey, K.C., Rayamajhi, A., Chowdhury, M., Bhavsar, P., Martin, J., 2016. Vehicle-to-vehicle (V2V) and vehicle-to-infrastructure (V2I) communication in a heterogeneous wireless network–Performance evaluation. Transportation Research Part C: Emerging Technologies, 68, 168-184.

Du, L., Dao, H., 2015. Information dissemination delay in vehicle-to-vehicle communication networks in a traffic stream. IEEE Transactions on Intelligent Transportation Systems, 16 (1), 66-80.

Du, L., Gong, S., Wang, L., Li, X., 2016. Information-traffic coupled cell transmission model for information spreading dynamics over vehicular ad hoc network on road segments. Transportation Research Part C: Emerging Technologies, 73, 30-48.

Gross, D., Shortle, J. F., Thompson, J. M., Harris, C. M., 2008. Fundamentals of queuing theory, 4th Edition. John Wiley & Sons.

Hisham, A., Ström, E. G., Brännström, F., Yan, L., 2017. Scheduling and power control for V2V broadcast communications with adjacent channel interference. arXiv preprint, arXiv:1708.02444.

Hisham, A., Sun, W., Ström, E. G., Brännström, F., 2016. Power control for broadcast V2V communications with adjacent carrier interference effects. IEEE International Conference on Communication, May, pp.1-6, Kuala Lumpur, Malaysia.

Kim, Y. H., Peeta, S., He, X., 2017. Modeling the information flow propagation wave under vehicle-to-vehicle communications. Transportation Research Part C: Emerging Technologies, 85, 377-395.

Kim, Y. H., Peeta, S., He, X., 2018. An analytical model to characterize the spatiotemporal propagation of information under vehicle-to-vehicle communications. IEEE Transactions on Intelligent Transportation Systems, 19 (1), 3-12.

Lighthill, M. J., Whitham, J. B., 1955. On kinematic waves. I: Flow movement in long river; II: A theory of traffic flow on long crowded roads. Proceedings of Royal Society A, 229 (1178), 281-345.

Li, M. Y., Graef, J. R., Wang, L., Karsai, J., 1999. Global dynamics of a SEIR model with varying total population size. Mathematical Biosciences, 160 (2), 191-213.

Li, M. Y., Muldowney, J. S., 1995. Global stability for the SEIR model in epidemiology. Mathematical biosciences, 125 (2), 155-164.

Noori, H., Olyaei, B. B., 2013. A novel study on beaconing for VANET-based vehicle to vehicle communication: Probability of beacon delivery in realistic large-scale urban area using 802.11p. 2013 International Conference on Smart Communications in Network Technologies, June, pp. 1-6, Paris, France.

Richards, P. I., 1956. Shock waves on the highway. Operations Research, 4, 42-51.

Smith, H. L., Wang, L., Li, M. Y., 2001. Global dynamics of an SEIR epidemic model with vertical transmission. SIAM Journal on Applied Mathematics, 62 (1), 58-69.

Talebpour, A., Mahmassani, H. S., Bustamante, F. E., 2016. Modeling driver behavior in a connected environment: Integrated microscopic simulation of traffic and mobile wireless telecommunication systems. Transportation Research Record: Journal of the Transportation Research Board, 2560, 75-86.

Wang, J., Kim, Y. H., He, X., Peeta, S., 2018. Analytical model for information flow propagation wave under an information relay control strategy in a congested vehicle-to-vehicle communication environment. Transportation Research Part C: Emerging Technologies, 94, 1-18.

Wang, X., 2007. Modeling the process of information relay through inter-vehicle communication. Transportation Research Part B: Methodological, 41 (6), 684-700.

Wang, X., Adams, T. M., Jin, W. L., Meng, Q., 2010. The process of information propagation in a traffic stream with a general vehicle headway: A revisit. Transportation Research Part C: Emerging Technologies, 18 (3), 367-375.

Wang, X., Yin, K., Zhang, Y., 2012. A Markov process for information propagation via inter-vehicle communication along two parallel roads. IEEE Transactions on Wireless Communication 11 (3), 865-868.

Wang, X., Yin, K., Qin, X., 2011. An approximate Bernoulli process in information propagation through inter-vehicle communication along two parallel roads. Transportation Research Part C: Emerging Technologies, 19 (3), 469-484.

Wang, X., Yin, K., Yan, X., 2015. Vehicle-to-vehicle connectivity on parallel roadways with large road separation. Transportation Research Part C: Emerging Technologies, 52, 93-101.

Zhang, J., Han, G., Qian, Y., 2016. Queuing theory based co-channel interference analysis approach for high-density wireless local area networks. Sensors, 16 (9), 1348.